# Analog of surface preroughening in a two-dimensional lattice Coulomb gas

Santi Prestipino*
*Istituto Nazionale per la Fisica della Materia, Unità di Ricerca di Messina, Messina, Italy*
*and Dipartimento di Fisica, Università degli Studi di Messina, Contrada Papardo, 98166 Messina, Italy*

Elaborating on previous theoretical treatments of the roughening transition, I provide the exact mapping of a surface model which undergoes both roughening and preroughening onto a mixture of unit and double charges living on the square lattice. Depending on the model parameters, the preroughening transition of the surface can be either continuous or discontinuous. Using the surface temperature as a control parameter, the dual Coulomb gas accordingly undergoes two consecutive phase transitions through equilibrium phases whose dielectric behavior is analyzed in terms of Monte Carlo simulation and exact finite-size calculations. Right at the preroughening point of the surface model, the charged mixture behaves like an electric insulator (provided preroughening is second order), while showing metallic behavior below and above that temperature, at least up to the roughening temperature where the conventional metal-insulator transition occurs.

## I. INTRODUCTION

Thermal disordering (i.e., roughening) of a crystal surface is usually described within the solid-on-solid (SOS) approximation. Two regimes are thus identified, separated by a critical value, $T_{\mathrm{R}}$, of the temperature $T$. For $T<T_{\mathrm{R}}$, the surface is smooth or flat, being pinned to a definite crystal plane; above $T_{\mathrm{R}}$, thermal excitations in the form of surface steps cause the delocalization of the surface which in turn freely wanders in space like a Gaussian interface. After the pioneering work of Chui and Weeks [1], roughening is believed, as a two-dimensional (2D) phase transition, to be in the Kosterlitz-Thouless (KT) universality class (together with the XY model, 2D melting, and the neutral lattice Coulomb gas, just to mention only a few 2D statistical models that exhibit this kind of criticality).

It was only in the late 1980s that Den Nijs and Rommelse provided a concrete example of a critical precursor of roughening, called preroughening (PR) [2,3]. Its promoting mechanism is an effective repulsion between *parallel* surface steps (i.e., both up or both down) which could shift roughening up in temperature, enough to unveil PR. Starting from the PR temperature $T_{\mathrm{PR}}$, the surface gets disordered just in the first layer, due to proliferation of up and down steps which, however, will retain a strict up-down order all the way up to $T_{\mathrm{R}}$. As a result, in the event of PR, the surface remains flat below $T_{\mathrm{R}}$, although its outermost layer is half-occupied for $T>T_{\mathrm{PR}}$ [the intermediate phase between PR and roughening is called the disordered-flat (DOF) phase].

As yet, the only unanimously recognized experimental realization of the preroughening-roughening scenario is found in the fcc(111) surface of rare-gas crystals [4]. In spite of this, PR still continues to arouse curiosity and theoretical interest for its amazing properties. For instance, Monte Carlo simulation [5], renormalization-group [6], and mean field theory [7] indicate that the PR point is an isolated point outside the rough phase where the interface width diverges in the infinite-size limit (at least, as long as PR is continuous). Moreover, it seems plausible that, in rare-gas surfaces, the (first-order) PR transition jointly occurs with the onset of surface melting [8]. Anyway, at present there is no clear evidence that the PR phenomenon is of any relevance to unreconstructed metal surfaces, although the experimental setup and/or long equilibration times in atomistic simulations could easily prevent the observation of a DOF-phase separation [9].

The present work somewhat deviates from the mainstream, in that it deals with a purely theoretical question. As mentioned above, Chui and Weeks were able to prove that roughening is an (infinite-order) KT phase transition by exactly mapping the discrete-Gaussian SOS model onto the neutral lattice Coulomb gas (CG) model. In particular, while the metallic phase of the CG model is in a correspondence with the smooth phase of the surface, the insulating phase is the counterpart of the rough phase. In this respect, the question naturally arises as to whether a system of charges exists which undergoes a further phase transition (other than the common metal-insulator one) having the features of PR. The answer, see the details in the Appendix, is affirmative. I was able to devise a SOS model with both PR and roughening which can be exactly mapped onto a version of the CG model. As a result, also the DOF phase of a surface will gain a precise translation into the language of electric charges.

The outline of this paper is the following. In Sec. II a suitably deformed sine-Gordon SOS model is introduced, for which the existence of PR and roughening is proved. Then, in Sec. III, the mapping of this model onto a CG model is developed, and the phase diagram of the latter is fully worked out. The structure of the various phases is mainly argued from the value of the dielectric constant and from the statistics of isolated and bound charges. Next, I give a summary of the main conclusions in Sec. IV.

## II. THE SOLID-ON-SOLID MODEL: PHASE DIAGRAM

### A. Motivation

In Ref. [1], the discrete-Gaussian (DG) SOS model and the 2D neutral lattice CG model are shown to be mutually dual, in that their partition functions are simply proportional to each other even though the temperature scales are inverted. In particular, surface roughening and the metal-insulator transition on a 2D lattice are nothing but different representations of the same critical phenomenon. The Hamiltonian of the DG model reads

$$H_{\rm DG}=J\sum_{\langle x,y\rangle}(h_x-h_y)^2, \tag{1}$$

for integer-valued heights defined on, say, the square lattice. The sum in Eq. (1) is over all distinct pairs of neighboring lattice sites and $J$ measures the positive cost of a step element. Both the site coordinates and the surface heights are hereafter taken to be dimensionless, i.e., given in terms of suitable in-plane and off-plane lattice constants, in such a way that $J$ sets the energy scale.

*Email address: santi.prestipino@unime.it

Following the lesson of Ref. [2], one could naively think to induce the stabilization of the DOF phase in the DG model by simply adding a suitable next-nearest-neighbor (NNN) Gaussian term to $H_{\rm DG}$. Although this term does not contrast step proliferation, it will nonetheless disfavor the appearance of nearby parallel steps, thus setting the stage for the DOF phase. More important, the amended Hamiltonian would be ready, apart from obvious modifications in the Chui-Weeks derivation, to be mapped onto a CG model. However, I checked by a series of extensive Monte Carlo runs that *no PR transition shows up in this way* [10]. As a matter of fact, the constraint $\Delta h=0,\pm1$ on the difference between nearest-neighbor (NN) heights, which is included in the (restricted-SOS) Hamiltonian of Ref. [2], proves to be a crucial ingredient for the DOF phase and there is apparently no easy way to handle this constraint analytically.

Luckily enough, another route is open to us. In a paper by Ohta and Kawasaki on a renormalization-group theory of the roughening transition [11], a different SOS Hamiltonian is being mapped onto a CG model:

$$H_{\rm OK}=J\sum_{\langle x,y\rangle}(h_x-h_y)^2-kT\sum_x \ln\left[1+\frac{y}{kT}\cos(2\pi h_x)\right], \tag{2}$$

where $k$ is the Boltzmann constant, $y$ is a positive energy, and the heights are *real* variables defined on the square lattice. At variance with Ref. [1], the partition function of $H_{\rm OK}$ is proportional to the partition function of the neutral square-lattice CG model with unit charges only and nonzero chemical potential. The new term on the right-hand side of Eq. (2) is a pinning term that keeps track of the crystal structure. The Hamiltonian (2), which reduces to a sine-Gordon Hamiltonian for small $y/(kT)$, should be regarded as a coarse-grained version of the "true" microscopic SOS Hamiltonian [somehow, each $h_x$ in Eq. (2) is an average of many microscopic heights]. In view of this, no surprise if $T$ explicitly appears in Eq. (2). From the viewpoint of renormalization-group theory, $H_{\rm OK}$ and the underlying SOS Hamiltonian would anyway exhibit the same thermodynamics, at least as far as the nature of phases and of the transitions between them are considered.

Insofar as one is interested in the CG analog of the DOF phase, the advantage of using a Hamiltonian of the sine-Gordon type soon becomes evident. In Ref. [7] a very accurate variational theory of both PR and roughening is generated using a modified sine-Gordon SOS Hamiltonian and a Gaussian *ansatz* for its free energy. Taking advantage of this information, we are rather spontaneously led to consider the following field theory, which is a modification of $H_{\rm OK}$:

$$H_{\rm SOS}=J\sum_{\langle x,y\rangle}(h_x-h_y)^2-\frac{1}{\beta}\sum_x \ln\left[\frac{1+\beta y_2\cos(2\pi h_x)+\beta y_4\cos(4\pi h_x)}{1+\beta y_2+\beta y_4}\right], \tag{3}$$

where $y_2$ and $y_4$ are suitable functions of the temperature and $\beta=1/(kT)$ (note that, for small values of $\beta y_2$ and $\beta y_4$, the sine-Gordon Hamiltonian of Ref. [7] is exactly recovered).

In order to assure that $H_{\rm SOS}$ shows PR at a given $T=T_{\rm PR}$ (prior to roughening), I tentatively set

$$y_2(T)=C(T_{\rm PR}-T)\quad\text{and}\quad y_4(T)>0, \tag{4}$$

with $C>0$. As discussed in Ref. [7], positive values of $y_2$ and $y_4$ will favor integer heights at low temperatures, hence a smooth surface below $T_{\rm PR}$. When $y_2<0$, heights would rather stay closer to a half-integer number, whence a different phase will be stable for $T>T_{\rm PR}$ where the first surface layer is only half-occupied (DOF phase). A PR transition will then occur at $T=T_{\rm PR}$. At high temperatures, entropic considerations will eventually prevail and the DOF phase will be overtaken in stability by the rough phase. We recall that, according to the mean-field theory, roughening of the sine-Gordon SOS model occurs at $T_{\rm R}^{\rm (MF)}=4J/(\pi k)$, whereas PR turns from second- to first-order when $T_{\rm PR}<J/(\pi k)$.

Below in this section, I provide conclusive evidence that $H_{\rm SOS}$ indeed undergoes, besides roughening, also a PR transition (either continuous or discontinuous, depending on $T_{\rm PR}$). Here, we just anticipate our main result (see proof in the Appendix): Hamiltonian (3) is exactly dual to a version of the CG model, see Eqs. (11)–(13) below, with unit and double charges only, whose study will be the subject of Sec. III.

### B. Monte Carlo analysis

In the following, dimensionless $T$ units are used, setting the reduced temperature equal to $t=kT/J$. Moreover, for later comparison with the mean-field results of Ref. [7], the study of $H_{\rm SOS}$ is specialized to a couple of choices for the parameters, namely,

$$\frac{C}{k}=0.5,\ \frac{kT_{\rm PR}}{J}=0.5,\quad\text{and}\quad\frac{y_4}{J}=0.1\rightarrow\text{model A}, \tag{5}$$

$$\frac{C}{k}=0.5,\ \frac{kT_{\rm PR}}{J}=0.25,\quad \text{and}\quad \frac{y_4}{J}=0.1\rightarrow \text{model B}. \quad (6)$$

According to the mean-field analysis of a closely related model system, model A would correspond to second-order PR at $t_{\rm PR}\equiv kT_{\rm PR}/J=0.5$, whereas model B would undergo a first-order PR transition at $t_{\rm PR}=0.25$.

Considering model A first, an important question to ask is about the argument of the logarithm in Eq. (3): it must be positive in order for $H_{\rm SOS}$ to be properly defined. Observe that, for the chosen $y_2$ and $y_4$, the quantity $1+\beta y_2+\beta y_4>0$ for any $t$. On the contrary, when $t<t_{\rm min}\equiv(21-8\sqrt{6})/10\simeq0.1404$, there is a range of $\cos(2\pi h)$ values where $1+\beta y_2\cos(2\pi h)+\beta y_4\cos(4\pi h)$ is *negative*. However, since the phase-diagram region that really matters lies well apart from $t_{\rm min}$, no serious drawback is attached to the lacking of $H_{\rm SOS}$ definition below $t_{\rm min}$.

I use standard Metropolis Monte Carlo (MC) to study $H_{\rm SOS}$. Square lattices $L\times L$ of three sizes are considered, namely $L=24,48$, and 72, with periodic boundary conditions (PBC) (I have also carried out a small number of runs for $L=96$). A MC move consists of updating the height at a randomly chosen lattice site by a random change in the interval $[-\delta h_{\rm max},\delta h_{\rm max}]$; then, the move is accepted or rejected according to the Metropolis rule. As usual, the value of $\delta h_{\rm max}$ is being adjusted during the run in such a way as to keep the acceptance ratio of MC moves as close to 50% as possible. After due equilibration of the sample, as many as $10^6$ sweeps are generated, each sweep consisting of one average attempt per site to change the local $h$. The relevant averages are updated every 10 sweeps.

Besides other quantities, I calculate the mean square height difference $\delta h^2=\langle(h_x-\overline{h})^2\rangle$ [with $\overline{h}=(1/N)\Sigma_x h_x$ and $N=L^2$], the order parameter $P=\langle\mathcal{P}\rangle$, where $\mathcal{P}=(1/N)|\Sigma_x \exp(i\pi h_x)|$, and the order-parameter susceptibility $\chi_P=N(\langle\mathcal{P}^2\rangle-\langle\mathcal{P}\rangle^2)$. As a note of caution, I observe that the above expression for $P$ is different from that usually assumed when the heights are integer [5]. Accordingly, nonzero values of both $P$ and $\delta h^2$ will be found not only in the smooth phase but also in the DOF phase. Right at the PR point, however, and so long as PR is critical, $P$ would vanish in the thermodynamic limit (due to a rough surface landscape), implying a dip in the finite-size $P$ and a peak for $\chi_P$ close to the PR temperature. Concurrently, $\delta h^2$ will blow up like $\ln L$. A less singular behavior will occur at PR in case of a first-order transition.

To corroborate our conclusions, the statistical uncertainties affecting the relevant averages are also evaluated. These are defined as root mean square deviations of statistical averages, once many independent estimates of these quantities (block averages) are given. In most cases, grouping MC states in blocks of $5\times10^4$ sweeps should be enough. However, particular care must be paid for the susceptibility, whose values could be correlated over much longer segments of MC trajectory (see below). Finally, errors are more severe close to second-order transition points, due to unlimited growth of decorrelation times.

In Fig. 1, MC simulation data are reported for model A. In the top panel I plot the average square height difference $\delta h^2$. The clearcut maximum at $t\simeq0.5$ is the most compelling evidence of a PR transition occurring, in the infinite-size limit, presumably right at $t_{\rm PR}$. Here, my data denounce a clear logarithmic increase of $\delta h^2$ with $L$, hence second-order PR (see the inset of Fig. 1 top). Below $t_{\rm PR}$, $\delta h^2$ appears to saturate and the same happens in the interval between $t_{\rm PR}$ and the roughening temperature $t_{\rm R}$, located somewhere around 1. At sufficiently high temperatures, my data indeed confirm that $\delta h^2$ increases logarithmically with $L$.

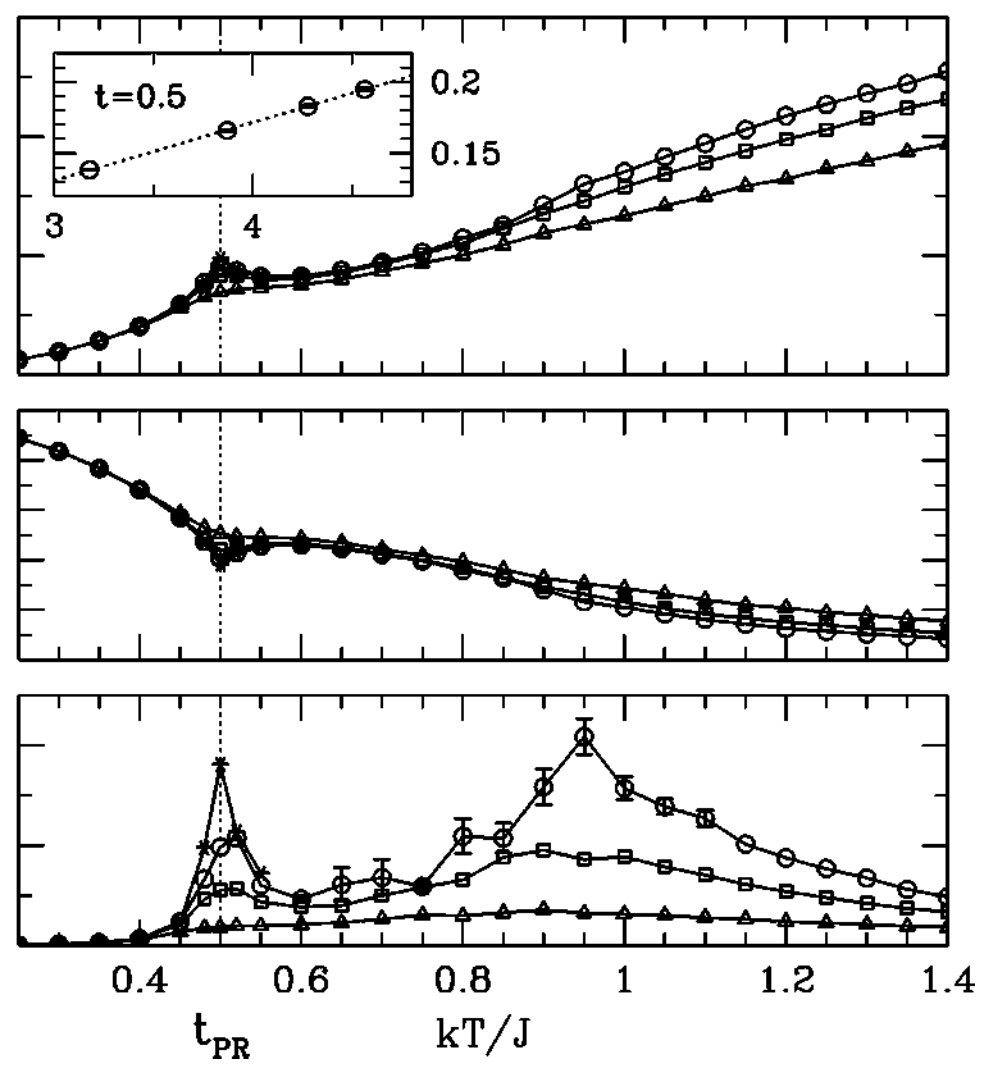

FIG. 1. MC simulation results for model A [Eqs. (3), (4), and (5)]. Data points relative to four lattice sizes, $L=24$ (△), 48 (□), 72 (○), and 96 (*), are plotted. Top: average square height difference $\delta h^2$. In the inset, finite-size scaling behavior of $\delta h^2$ at $t=t_{\rm PR}$, plotted as a function of $\ln L$ (circles, MC data; dotted line, best linear fit, $\delta h^2=0.008\,892\,33+0.040\,631\,8\ln L$). The error bars are also shown. Straight line segments between the points are drawn as a guide to the eye. Center: the order parameter $P$. As for $\delta h^2$, the statistical errors for $P$ are negligible, smaller than the size of the symbols. Bottom: the order-parameter susceptibility $\chi_P$ (the error bars are estimated from block averaging). The reported data clearly indicate the existence of a PR phase transition close to, if not exactly at, $t=0.5$. It is much more difficult to say where roughening occurs, presumably near $t=1$.

The order parameter $P$ is plotted in Fig. 1 (center). The expected dip at $t_{\rm PR}$ as well as the vanishing at $t_{\rm R}$ are both evident. Near each transition point, the order-parameter susceptibility $\chi_P$ shows a peak whose height increases with the system size [see Fig. 1 (bottom)]. Furthermore, the specific heat remains finite at any $t$ (not shown), suggesting a negative PR specific-heat exponent.

A picture of how the surface looks in the various phases can be drawn from the MC-time evolution of the current mean surface height $\overline{h}$. In Fig. 2, this evolution is shown for $L=72$, at four distinct temperature values: $t=0.45$ (smooth phase), $t=0.5$ (PR), $t=0.7$ (DOF phase), and $t=1.2$ (rough phase). We clearly see that $\overline{h}$ is roughly constant in both flat phases. In fact, it carries out only small fluctuations around an integer or an integer-plus-one-half value according to whether the phase is smooth ($t$=0.45) or DOF ($t$=0.7). Occasionally, $\overline{h}$ performs enormous "jumps" of unit length (giant fluctuations of the SOS interface as a whole), which are more frequent the smaller the surface size. Each jump simply indicates a transition (during the simulation run) between two different, although perfectly equivalent, realizations of the same phase. This behavior closely resembles the one originally discovered in the FCSOS model [5]. Such jumps occur very rapidly on the time scale of the simulation. Moreover, each time a jump occurs, it causes an abrupt variation of $\chi_P$ which, when the total number of jumps during the run is very small, makes less reliable the statistical error of $\chi_P$ as calculated through block averaging.

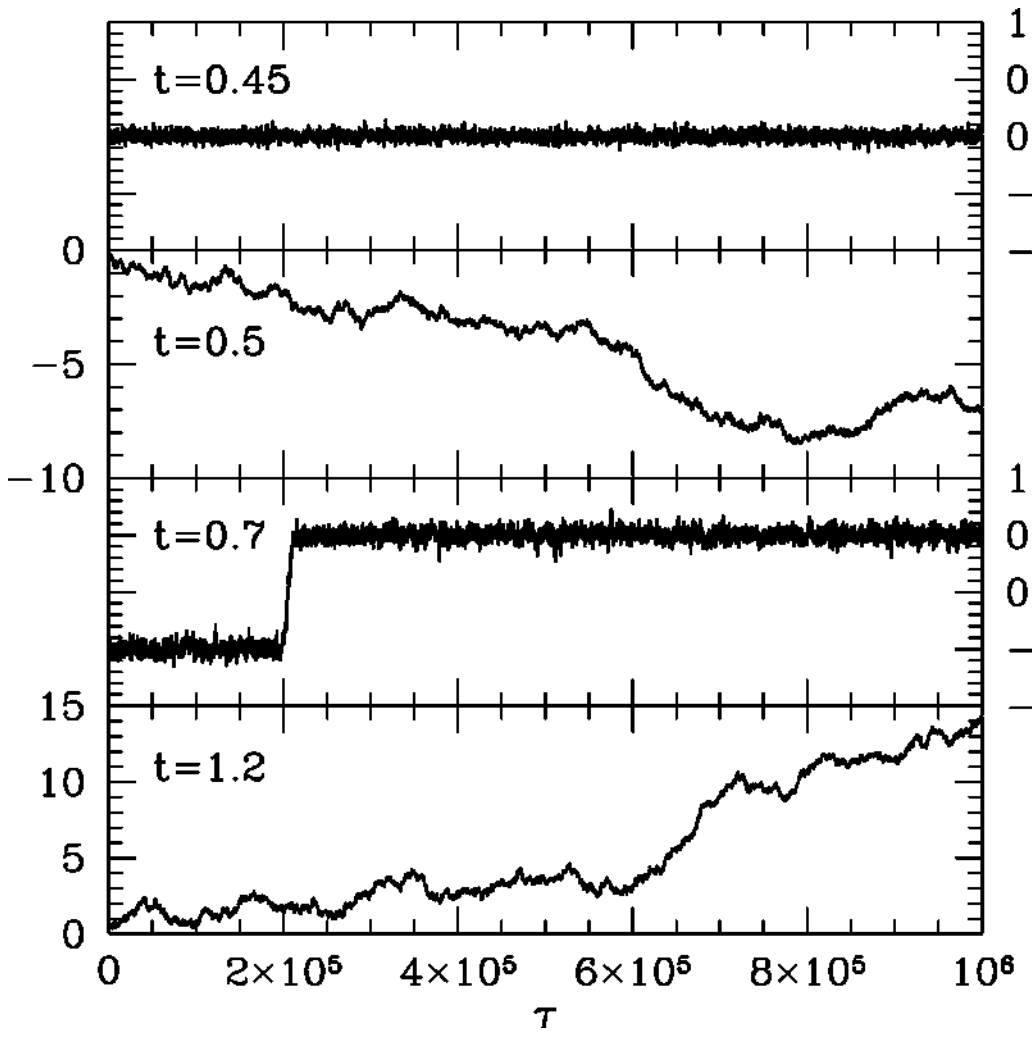

FIG. 2. Run-time evolution of the mean surface height $\overline{h}$ of a $72\times72$ system (model A), for a number of (reduced) temperatures: from top to bottom, $t=0.45$ (smooth phase), $t=0.5$ (PR), $t=0.7$ (DOF phase), and $t=1.2$ (rough phase). $\tau$ is the MC time as measured in sweeps. Rather evident is the rupture, in the behavior of $\overline{h}$, that occurs at PR. Moreover, the behavior at PR is similar to that in the rough phase. The unit jump of $\overline{h}$ at $t=0.7$ is a finite-size effect.

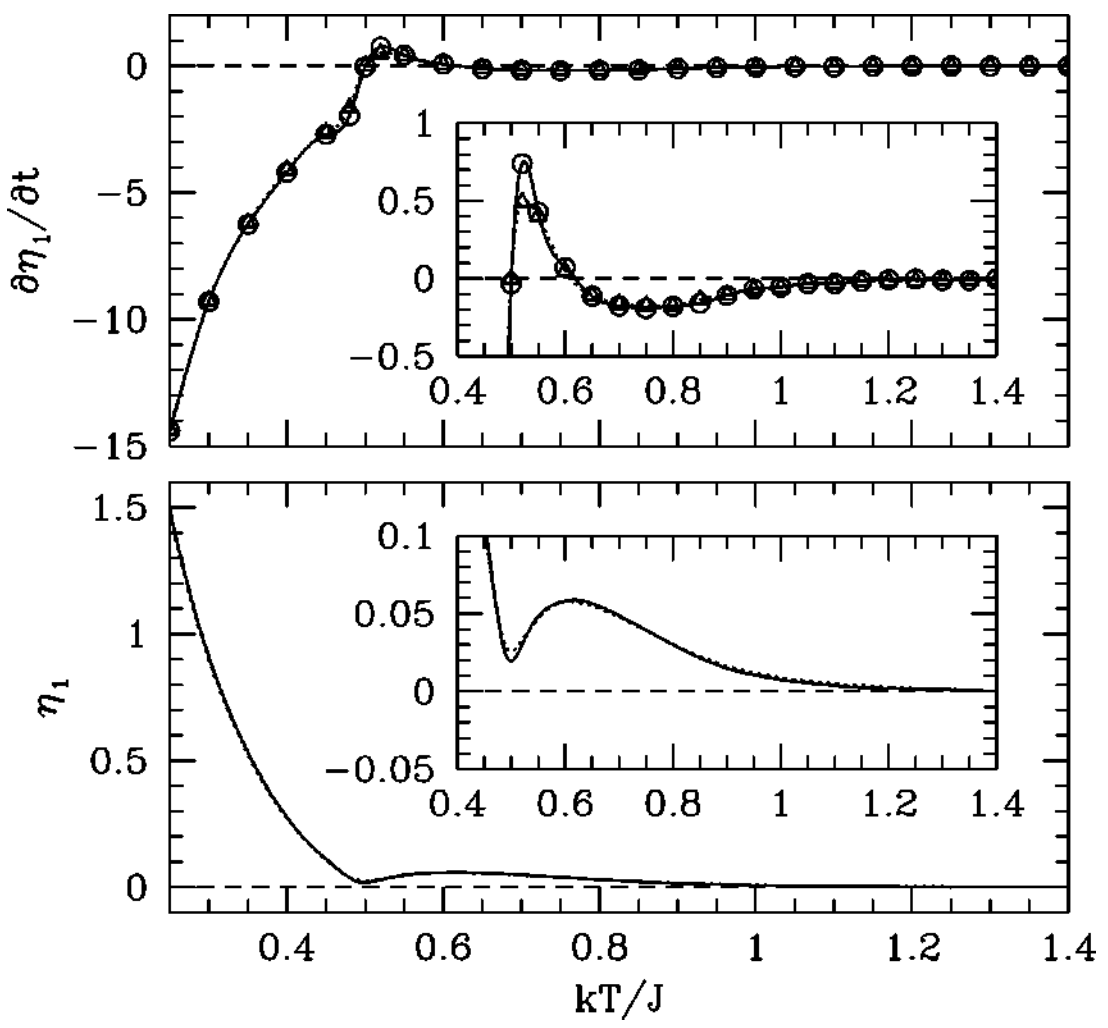

FIG. 3. MC simulation results for model A [Eqs. (3), (4), and (5)]. Data points for square lattices of two sizes, $L=48$ (△) and 72 (○), are plotted. Above: thermal derivative of the tilting free energy as a function of $t$, at fixed $C$, $t_{\rm PR}$, and $y_4$. The lines through the data points are spline functions (dotted line, $L=48$; continuous line, $L=72$). Below: the tilting free energy. On account of these results, we conclude that, in the infinite-size limit, $\eta_1=0$ at PR, while keeping a nonzero value in both flat phases as a result of the rigidity of the surface towards tilting. On approaching the rough phase, $\eta_1$ gradually lowers, eventually vanishing at the roughening temperature.

A further, independent evidence of the DOF phase comes from a measurement of the free-energy cost for tilting the surface. This is defined as $\eta_1=L\beta(f_1-f_0)$, where $f_0$ is the free energy per site when full PBC are applied and $f_1$ is the same quantity for the tilted surface. In order to tilt the surface, the use of proper "periodic-step boundary conditions" is mandatory. Unfortunately, the standard method for $\eta_1$ which uses the transfer matrix [2,12] cannot be applied in this case since the heights are real numbers. In spite of this, at least the thermal derivative of $\eta_1$ can be obtained in a MC experiment by inserting a unit step at the vertical boundaries of the simulation box and recording the resulting change in a number of averages.

Starting from the relation $-\beta F=\ln Z$, a rather lenghty calculation first yields

$$\left(\frac{\partial\beta F}{\partial\beta J}\right)_{C,t_{\rm PR},y_4}=\left\langle\sum_{\langle x,y\rangle}(h_x-h_y)^2\right\rangle+(1+\beta y_2+\beta y_4)^{-1}\times\left\langle\sum_x\frac{\frac{C}{k}t_{\rm PR}[1-\cos(2\pi h_x)]+\frac{y_4}{J}[1-\cos(4\pi h_x)]+\frac{C}{k}\frac{y_4}{J}[\cos(4\pi h_x)-\cos(2\pi h_x)]}{1+\beta y_2\cos(2\pi h_x)+\beta y_4\cos(4\pi h_x)}\right\rangle. \quad (7)$$

In deriving this result, the $y_2$ dependence on $\beta J$ has been considered too. From the definition of $\eta_1$, we then have

$$\frac{\partial(\eta_1/L)}{\partial\beta J}=\frac{1}{L^2}\left(\frac{\partial\beta F_1}{\partial\beta J}-\frac{\partial\beta F_0}{\partial\beta J}\right), \quad (8)$$

where each derivative is being performed at fixed $C,t_{\rm PR}$, and $y_4$.

I expect that, by its very definition, $\eta_1$ is nonzero in any flat phase, while vanishing like $L^{-1}$ at a second-order PR transition and in the whole rough phase [2]. In a *finite* sys-tem, $\eta_1$ would be minimum at a temperature $t^*$ close to $t_{\rm PR}$, hence the thermal derivative of $\eta_1$ will be negative for $t<t^*$ and positive for $t>t^*$. Indeed, this is what I observe (see Fig. 3).

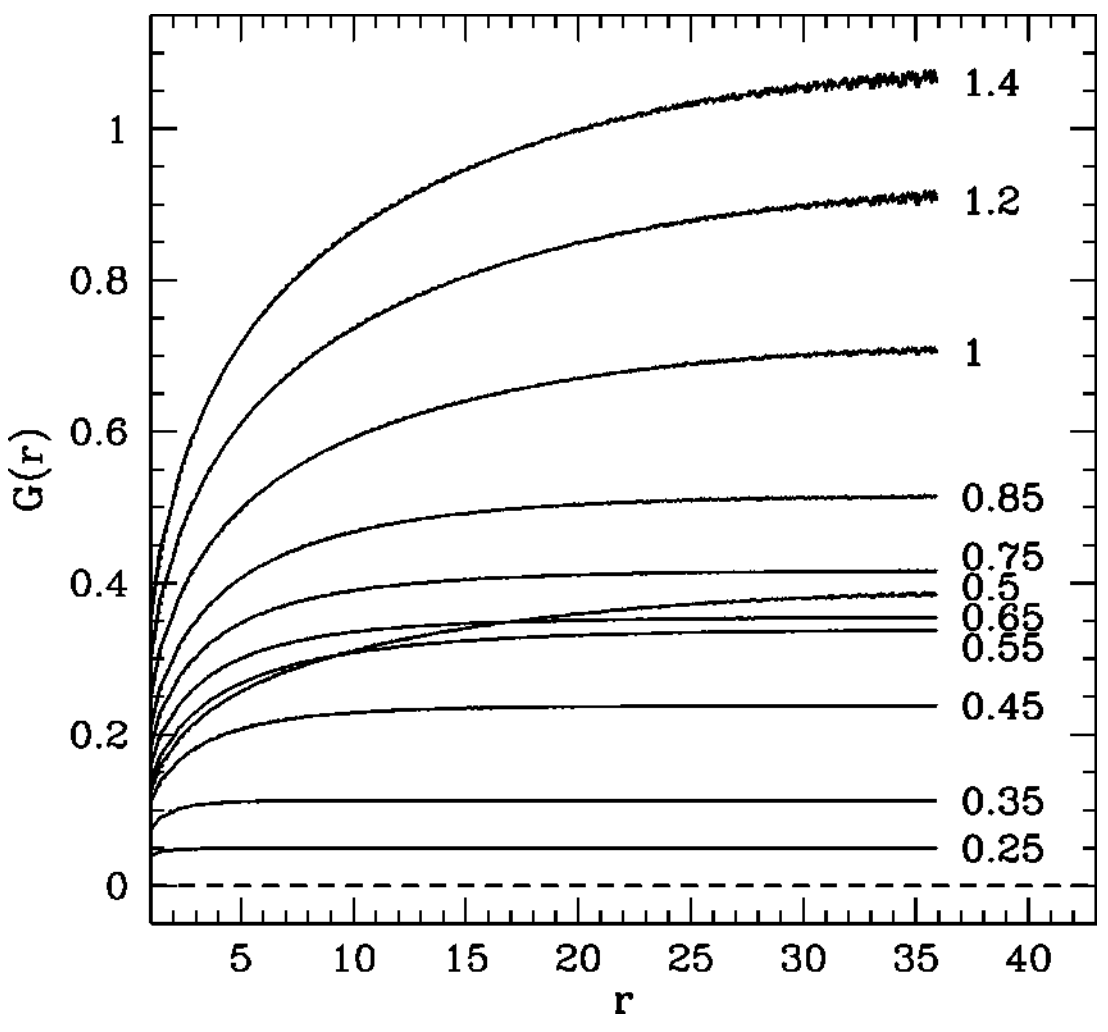

FIG. 4. Radial distribution function $G(|x|)=\langle (h_x-h_0)^2\rangle$ for the heights (model A). The data points are for a $72\times 72$ lattice and a number of temperatures $t$ (as indicated beside each line). The dotted lines are the best-fitting curves according to Eq. (10) (in the plot they are almost indistinguishable from the MC profiles). Observe the difference between the trend of $G(r)$ to saturation, typical of a flat surface, and the logarithmic increase at large distance, which applies at PR and in the whole rough phase.

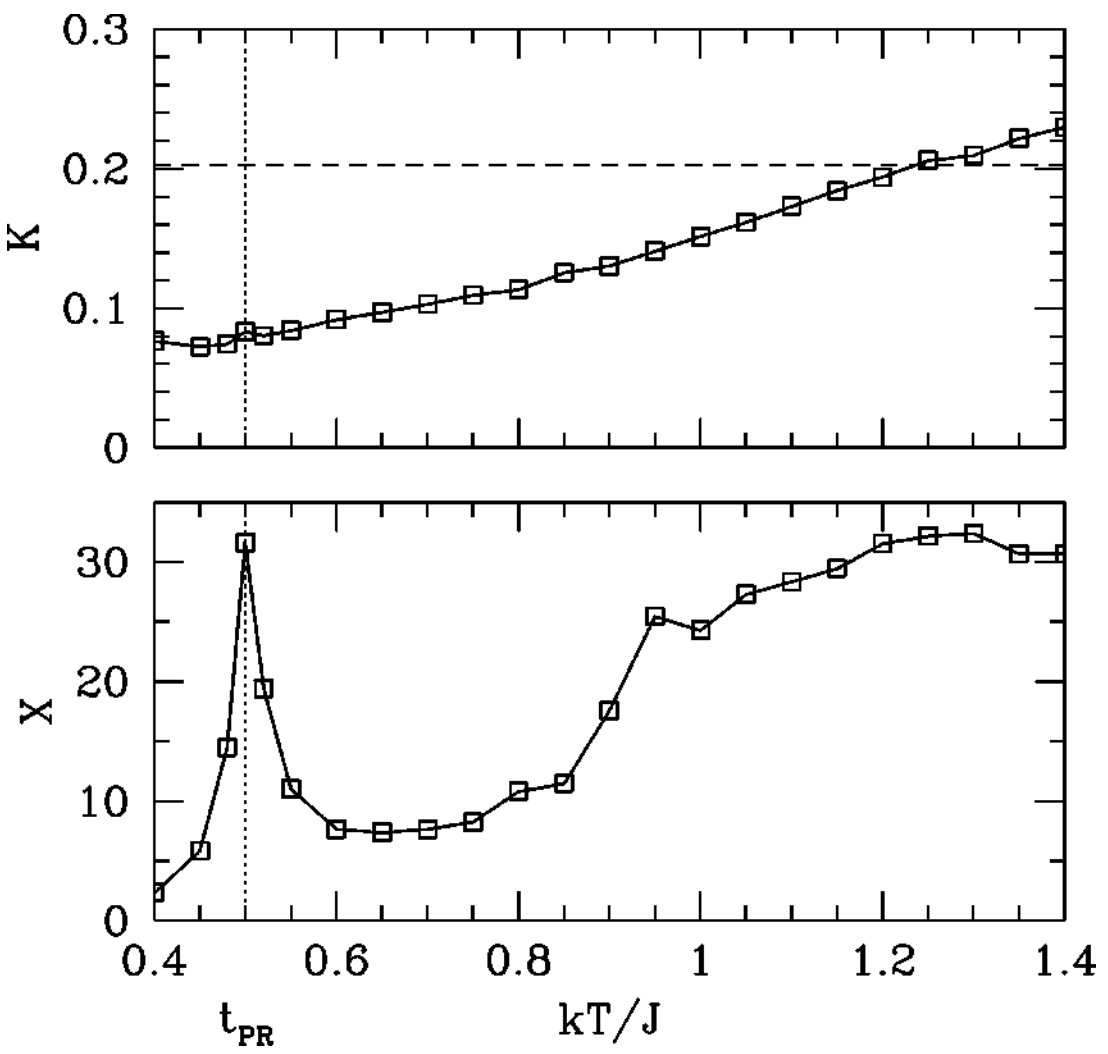

FIG. 5. Thermal evolution of the fitting parameters $K$ and $X$ for the same $72\times 72$ systems as in Fig. 4. $K$ is the roughness strength, which should take at roughening the universal value $2/\pi^2$ (dashed horizontal line, being crossed for $t\simeq 1.25$) and a smaller value at PR (here, $K\simeq 0.08$). $X$ is a sort of correlation length which, in the thermodynamic limit, would diverge at PR and in the whole rough phase.

Assuming $\eta_1$ to be exactly zero at our maximum simulation temperature $t=1.4$, the quantity plotted in Fig. 3 above can be further integrated to give the $\eta_1$ profile, shown in Fig. 3 below for the two sizes $L=48$ and $L=72$.

In the thermodynamic limit, $\eta_1$ is positive at low temperature, first vanishing at $t_{\rm PR}$. On heating, $\eta_1$ recovers a positive value in the DOF phase, until it vanishes again and for good when reaching the roughening temperature $t_{\rm R}$. As a result, the surface first delocalizes at $t_{\rm PR}$, but, due to a nonzero $\eta_1$ value between $t_{\rm PR}$ and $t_{\rm R}$, it remains flat in the DOF phase. It is necessary to wait until $t_{\rm R}$ for the surface to become delocalized again, now permanently. I note that the behavior of $\eta_1(t)$ just described is the same as suggested by the variational theory of Ref. [7] for the step free energy.

Another interesting subject is that of height-height correlations. Their character is studied in detail for $L=72$. In Fig. 4, I plot the correlation function

$$G(|x|)=\langle (h_x-h_0)^2\rangle, \tag{9}$$

for a number of $t$ values in the relevant temperature range. Because of the PBC, the $G$ profile levels off at $L/2$. This behavior is interpolated through the best (least-squares) logarithmic fit, as drawn from Refs. [13] and [14]:

$$G(|x|)\simeq -\frac{K(T)}{2}\ln(|x|^{-2}+X(T)^{-2})+C(T). \tag{10}$$

In fact, in order to extract the correlation length $\xi$, a more natural choice would have been to imagine an exponential damping for the large-distance profile of $G(|x|)$. However, the logarithmic fit proves to be more effective than the exponential fit in the *whole* range of $|x|$ values. Moreover, information about the location of phase transitions are equally accurate from the fit (10). In fact, although $X$ is not the correlation length, $X$ and $\xi$ behave similarly at any temperature, being both finite or both infinite. In particular, each transition point is characterized by a very large $X$ value, as demonstrated in Fig. 5. Hence, while $G(|x|)$ asymptotically saturates (to approximately $2\,\delta h^2$) when the surface is macroscopically flat, it shows a large-distance logarithmic increase at PR and in the whole rough phase.

Although not completely reliable, due to the finite system size and to the intrinsic inaccuracy of formula (10), the interface roughness $K$ attains the universal $2/\pi^2$ roughening value at $t\simeq 1.2$. Moreover, the $K$ value at PR is smaller than $2/\pi^2$, as indeed expected [7].

In closing, the question is addressed as to whether a different $t_{\rm PR}$ value would lead to a first-order PR (note that the location of roughening should not be significantly affected by a modification of $t_{\rm PR}$). For the choice pertaining to model B, $t$ should not fall below 0.1064 . . . for the logarithm in Eq. (3) to be defined.

In Fig. 6, some MC results for model B are compared with those relative to model A. PR is much likely first-order now, as being signaled by the leveling off of $P(t_{\rm PR})$ at a nonzero minimum value, by the apparently infinite jump of $\chi_P$ in the thermodynamic limit, and by the convergence of $\delta h^2$ to a finite value (see Fig. 6 inset, where the value of $\delta h^2$ at $t=0.25$ is plotted versus $\ln L$). This is also confirmed by the run-time evolution of $\overline{h}$ at $t=0.25$ (Fig. 7), which is radically different from the one we expect for a rough surface (that is, for second-order PR). Here, surface states with integer $\overline{h}$ are sampled within the same simulation run together with states where $\overline{h}$ is half-integer: this is a typical MC feature for thermodynamic phase coexistence which is also the reason for the blowing up of $\chi_P$ right at $t_{\rm PR}$.

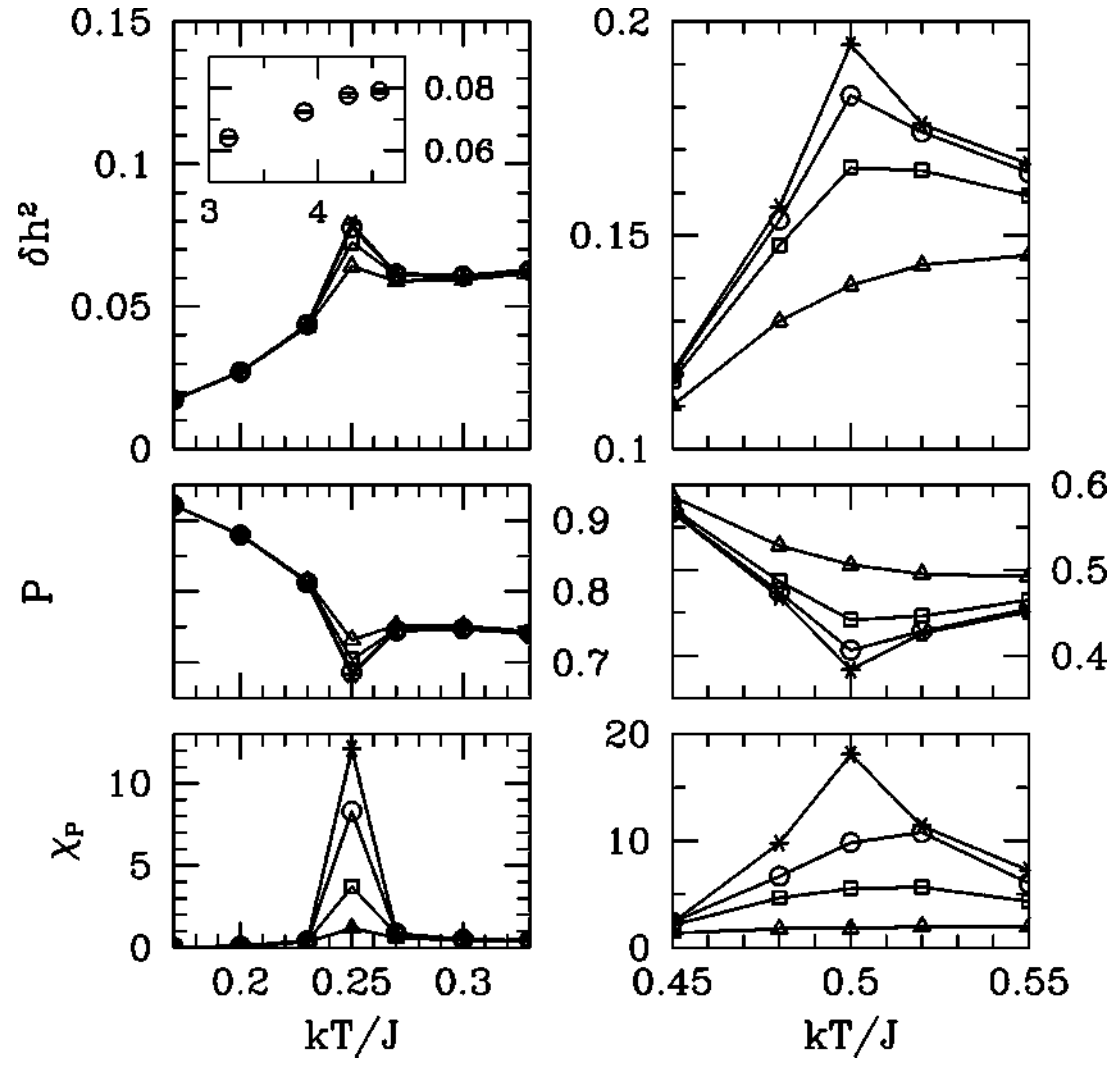

FIG. 6. MC data for the SOS model [Eqs. (3) and (4)]: comparison between model B ($t_{\rm PR}=0.25$, left) and A ($t_{\rm PR}=0.5$, right) (same sizes and notations as in Fig. 1). While PR is second-order for model A, it is most likely first-order for model B. In the inset, $\delta h^2$ vs $\ln L$ at $t=t_{\rm PR}=0.25$ (the error bars are also shown).

To conclude, my SOS model shows a stable DOF phase in the temperature range between $t_{\rm PR}$ and $t_{\rm R}\approx 1$. In close agreement with the mean-field prediction, the PR transition changes from second-order to first-order upon reducing the value of $t_{\rm PR}$. In the next section, I move on to study the CG model that is dual to $H_{\rm SOS}$.

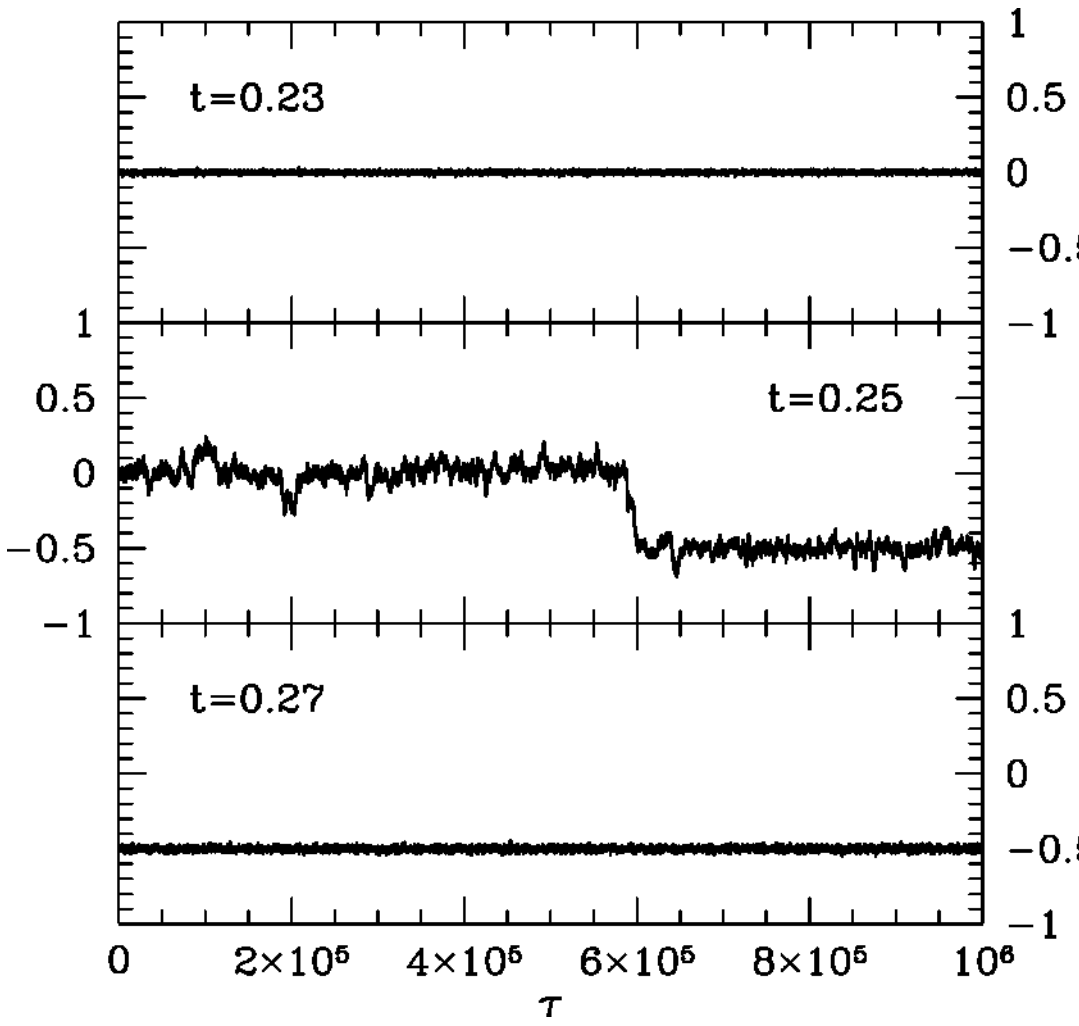

FIG. 7. Run-time evolution of the mean surface height $\overline{h}$ of a 96×96 system (model B), at three different temperatures: from top to bottom, $t=0.23$ (smooth phase), $t=0.25$ (PR), and $t=0.27$ (DOF phase). $\tau$ is the MC time as measured in sweeps. The half-unit jump of $\overline{h}$ at $t=0.25$ is the evidence that the smooth and the DOF phases do coexist, implying a first-order PR transition.

## III. THE DUAL COULOMB GAS: PHASE DIAGRAM

### A. The model

Now that I ascertained the existence of a stable DOF phase in the SOS model, I am in a position of unveiling the exact counterpart, in the CG language, of the DOF phase. As I prove in the Appendix, there is a CG model which is exactly dual to $H_{\rm SOS}$, implying three phases and two phase transitions for both. In particular, the transition points of the CG model fall at the same values of $t$ where the PR and roughening of the surface model occur.

The partition function of the CG model is given by

$$\Xi_{\rm Q}=\sum_{\{q_x\}}\delta_{\sum_x q_x,0}\exp[\beta_{\rm Q}\mu_1(N_1+N_{-1})+\beta_{\rm Q}\mu_2(N_2+N_{-2})]\exp\left(-\beta_{\rm Q}\sum_{x<y}V_{x,y}q_xq_y\right), \tag{11}$$

where $q_x=0,\pm 1,\pm 2$ is the charge at the $x$ position in the lattice and $N_q=\Sigma_x\delta_{q_x,q}$ is the current number of $q$ charges. Due to the Kronecker delta in Eq. (11), only those charge configurations that satisfy the neutrality condition $\Sigma_x q_x=0$ are to be summed over.

The square-lattice Coulomb potential reads:

$$V_{x,y}=\frac{\pi}{N}\sum_{p\neq 0}\frac{\exp[ip(x-y)]-1}{2-\cos p_x-\cos p_y}\sim-\ln(|x-y|)-\frac{1}{2}\ln 8-\gamma, \tag{12}$$

where $p=p_x\hat{x}+p_y\hat{y}$ is the dimensionless vector with components $p_x=2\pi m_x/L$ and $p_y=2\pi m_y/L$ (for $m_x,m_y=0,1,\ldots,L-1$). In Eq. (12), the large-distance behavior of $V$ is also indicated ($\gamma\simeq 0.577$ is the Eulero-Mascheroni constant). In Fig. 8, $V$ is plotted for a periodically repeated $24\times 24$ lattice as a function of the discretized distance from a reference position. When the minimum-image-distance convention is adopted, $V$ reaches its maximum absolute value at half of the box length. Therefore the large-distance logarithmic behavior only applies in the window (if it even exists at all) $1\ll r\ll L/2$. I wish to emphasize that, due to the discreteness of the embedding space, the translationally invariant lattice Coulomb potential is not spherically symmetric (for instance, the values of $V_{x,y}$ at $x-y=(0,5)$ and $x-y=(3,4)$ are slightly different, see Fig. 8, where two separate symbols appear for $r=5$).

Finally, the temperature and chemical potentials of the charges are well-defined functions of the reduced SOS temperature, also through the phenomenological parameters $y_2$ and $y_4$, given by

$$\beta_Q=\frac{\pi}{\beta J},\quad z_1\equiv\exp(\beta_Q\mu_1)=\frac{\beta|y_2|}{2},$$

$$\text{and}\quad z_2\equiv\exp(\beta_Q\mu_2)=\frac{\beta y_4}{2}. \tag{13}$$

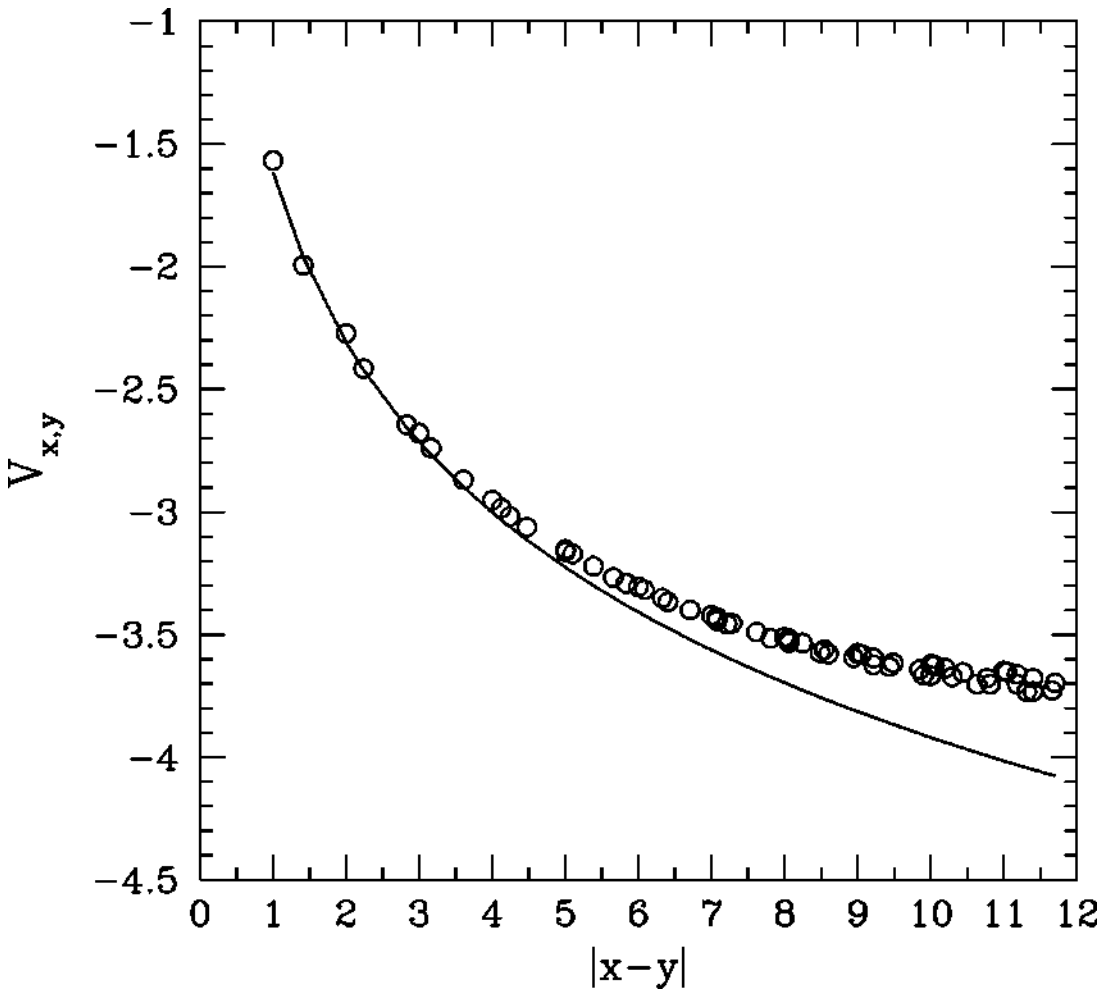

FIG. 8. The Coulomb potential $V$ for $L=24$, plotted as a function of the distance from a reference lattice point, up to $L/2$. The continuous line is the asymptotic behavior of $V_{x,y}$, that is, $-\ln|x-y|-(1/2)\ln 8-\gamma$ ($\gamma\simeq 0.577$), which, however, would only apply in the range $1\ll|x-y|\ll L/2$.

Note that the temperature $T_Q=1/\beta_Q$ of the charges is inversely proportional to the SOS temperature $t$ (this is exactly what the word duality is for). It should be observed that $T_Q,\mu_1$, and $\mu_2$ are all dimensionless quantities. Unless otherwise specified, I use for $y_2$ and $y_4$ the same expressions relative to model A. In this case, the quantities in Eq. (13) are plotted in Fig. 9 as a function of the SOS temperature $t$. By looking at this picture, we immediately realize that, in contrast to $\mu_2$, the chemical potential $\mu_1$ of unit charges has a nonmonotonic trend which, as we shall see, is solely responsible for the phase transition of the CG model at $t_{PR}$. Moreover, $\mu_1$ and $\mu_2$ are largely negative, which implies low values for the charge densities but for sufficiently high $T_Q$.

The argument $-(1/T_Q)\Sigma_{x<y}V_{x,y}q_xq_y$ of the exponential in Eq. (11) is negative for opposite charges. When $T_Q$ is low, the few charges present are preferentially bound together in neutral NN pairs (dipoles). On the contrary, charges of equal sign push each other away. Besides isolated double charges, also 2-dipoles (i.e., dipoles formed by double charges) are strongly suppressed at low temperature since the argument of the exponential is four times more negative than for 1-dipoles (for low $T_Q$, also the fugacity $z_2$ of double charges is smaller than the fugacity $z_1$ of unit charges).

A thorough study of model (11)–(13) necessarily implies the use of numerical simulation. However, before going on to illustrate the simulation procedure, I shall take advantage of the negative values attained by $\mu_1$ and $\mu_2$ in the relevant $t$ range for carrying out a perturbative study of the CG model. My intuition that the charged system under consideration is indeed a very dilute one (with the only exception of low $t$ values) will be consistently tested in a $12\times 12$ lattice through a truncated expansion of the partition function of the charges in powers of $z_1$ and $z_2$. Besides virtually *exact* reference results, this expansion will give us the possibility to check the correctness of the MC procedure.

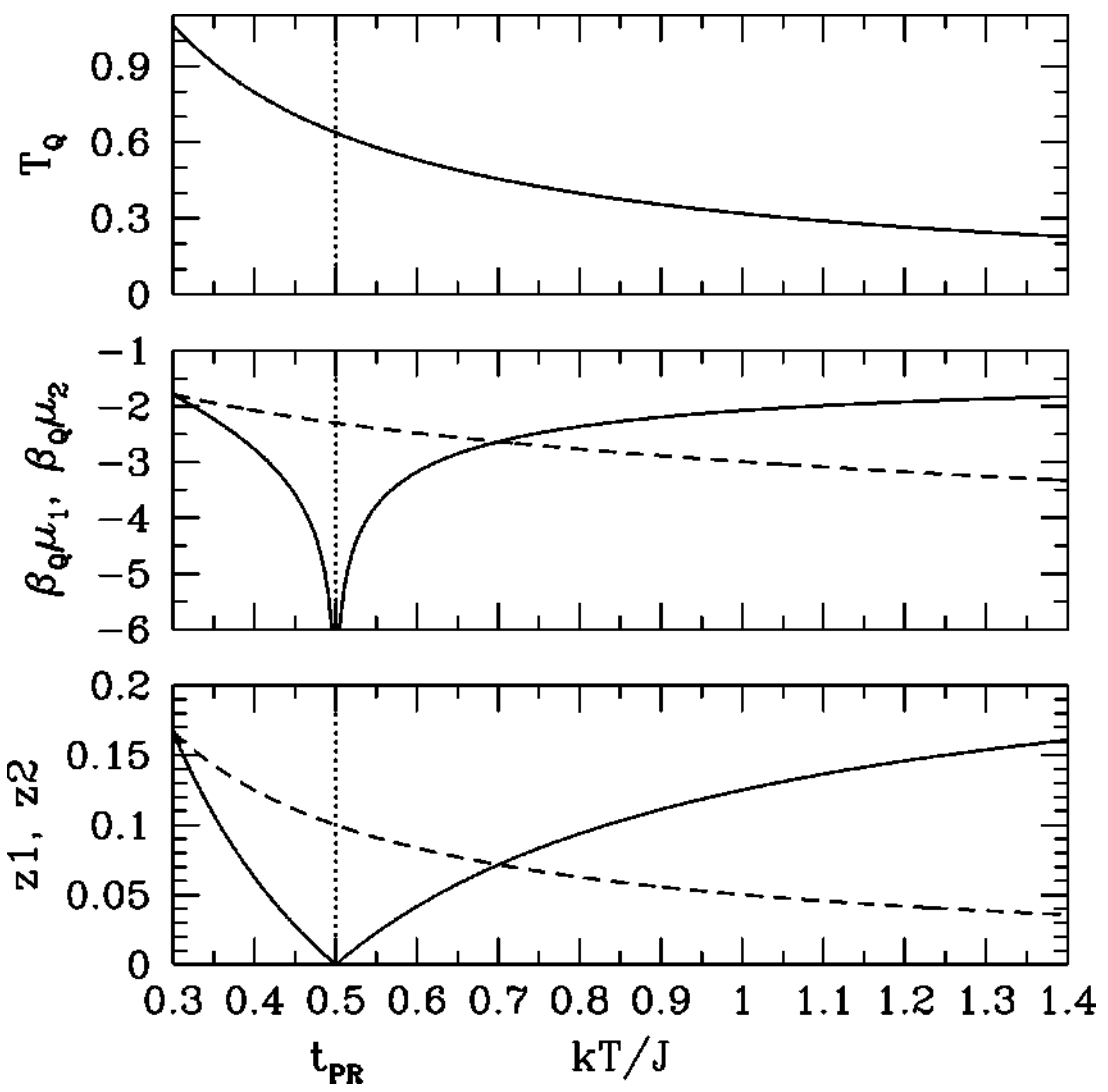

FIG. 9. The figure shows the evolution, as a function of the SOS temperature $t$, of the parameters (13) controlling the statistical mechanics of the CG model that is dual to the SOS model A. Top: temperature $T_Q=1/\beta_Q$ of the charges. Center: reduced chemical potential of unit ($\beta_Q\mu_1$, continuous line) and double charges ($\beta_Q\mu_2$, dashed line). Bottom: fugacity of unit ($z_1$) and double charges ($z_2$). Note the nonmonotonic trend of $\mu_1$ as a function of $t$ or $T_Q$, and its logarithmic singularity at $t_{PR}$.

Compared to the more conventional (Chui-Weeks) situation, a complication in my case arises because the insulating/rough phase is separated from the metallic/smooth phase by a further DOF phase. This is why any method to investigate in depth the statistical mechanics of a diluted system of charges, however small it might be, is welcomed as an invaluable opportunity.

### B. Low-fugacity expansion of the partition function

In Ref. [19] a truncated fugacity expansion of the partition function of a neutral square-lattice CG of unit charges was considered in order to explore a region of the phase diagram that MC simulation could not reach. In particular, all possible configurations of 2, 4, and 6 charges were enumerated. In that case, however, the positive value of $\mu$ severely restricted the validity of the expansion to relatively small temperature values.

Here, I attempt a similar expansion but for a system of *four* different species of charges. The grand-canonical partition function $\Xi_Q$ is written as

$$\Xi_Q=\sum_{N_1,N_{-1},N_2,N_{-2}\geqslant 0}\delta_{\sum_q qN_q,0}\,z_1^{N_1+N_{-1}}z_2^{N_2+N_{-2}}$$

$$\times Z_{N_1,N_{-1},N_2,N_{-2}}, \tag{14}$$

where $Z_{N_1,N_{-1},N_2,N_{-2}}$ (a function of $\beta_Q$ only) is the *canonical* partition function of a system containing a fixed number $N_q$ of $q$ charges, for $q=\pm 1,\pm 2$.

For given numbers $N_q$ of charges, the exact computation of $Z_{N_1,N_{-1},N_2,N_{-2}}$ requires one to sum a maximum number $N(N-1)(N-2)\cdots(N-N_{ch}+1)$ of Boltzmann weights, where $N_{ch}=N_1+N_{-1}+N_2+N_{-2}$ is the total number of charges on the lattice (in many cases, however, the use of symmetry arguments considerably simplifies the calculation). Moreover, the number of 4-tuples of non-negative integers $N_q$ satisfying $N_1-N_{-1}+2N_2-2N_{-2}=0$ rapidly grows with $N_{ch}$. Therefore it is clear that $\Xi_Q$ can be evaluated only when $L=\sqrt{N}$ is small and Eq. (14) is truncated to low order.

If only terms up to $N_{ch}=6$ are kept in Eq. (14), the partition function will read

$$\begin{aligned}\Xi_Q \simeq & 1+z_1^2Z_{1100}+z_2^2Z_{0011}+z_1^2z_2(Z_{2001}+Z_{0210})\\ &+z_1^2z_2^2Z_{1111}+z_1^4Z_{2200}+z_2^4Z_{0022}\\ &+z_1^4z_2(Z_{3101}+Z_{1310})+z_1^2z_2^3(Z_{2012}+Z_{0221})\\ &+z_1^4z_2^2(Z_{4002}+Z_{0420})+z_1^6Z_{3300}\\ &+z_1^4z_2^2Z_{2211}+z_1^2z_2^4Z_{1122}+z_2^6Z_{0033}, \qquad (15)\end{aligned}$$

where, due to charge-inversion symmetry of the Hamiltonian, $Z_{2001}=Z_{0210}$, and so on. Within the same multiple loop that gives $Z_{N_1,N_{-1},N_2,N_{-2}}$ as an output, I also evaluate the canonical average of the system energy $H_{N_1,N_{-1},N_2,N_{-2}}$. With this information at hand, I readily calculate the following grand-canonical averages: (1) the number densities $\rho_q=(1/N)\langle N_q\rangle$ (where $\rho_1=\rho_{-1}$ and $\rho_2=\rho_{-2}$ owing to the fact that the chemical potential is the same for opposite charges); (2) the energy per site $u=(1/N)\langle\Sigma_{x<y}V_{x,y}q_xq_y\rangle$; and (3) the number histogram. Only at this moment, the reliability of the truncation (15) can be judged on the basis of the calculated probability density for $N_{ch}$. For a $12\times 12$ lattice and $t_{PR}=0.5$ (corresponding to model A), the probability of having more than six charges on the lattice is absolutely negligible when $t>0.45$ (see Fig. 10). This is enough for considering as *virtually exact* for $t>0.45$ any result obtained from approximating the partition function of the $12\times 12$ system through Eq. (15) (observe that the relevant region of the CG model phase diagram lies entirely within this $t$ interval).

For reasons which will become clear in a moment, I am also interested in the spatial correlations between the charges. I calculate the standard two-point correlation functions $\langle c_{xq}c_{yq'}\rangle\equiv\rho_q\rho_{q'}g_{qq'}(x,y)$ (for $x\neq y$), where $c_{xq}$ is an occupation number ($c_{xq}=1$ if site $x$ is occupied by a $q$ charge, 0 otherwise). For any of the allowed 4-tuples $\{N_q\}$ [see Eq. (15), where all of the 13 cases are listed], there is a (possibly zero) contribution to $\langle c_{xq}c_{yq'}\rangle$; in turn, any such contribution, suitably weighted by some powers of the fugacities, will concur to fixing $g_{qq'}$ (a grand-canonical average). Whenever applicable, use of symmetry considerations allows one to speed up the loop calculations, hence to reduce the computational effort. Likewise the potential $g_{qq'}(x,y)$ is not generally a function of $|x-y|$ only, although deviations from radial symmetry are minute.

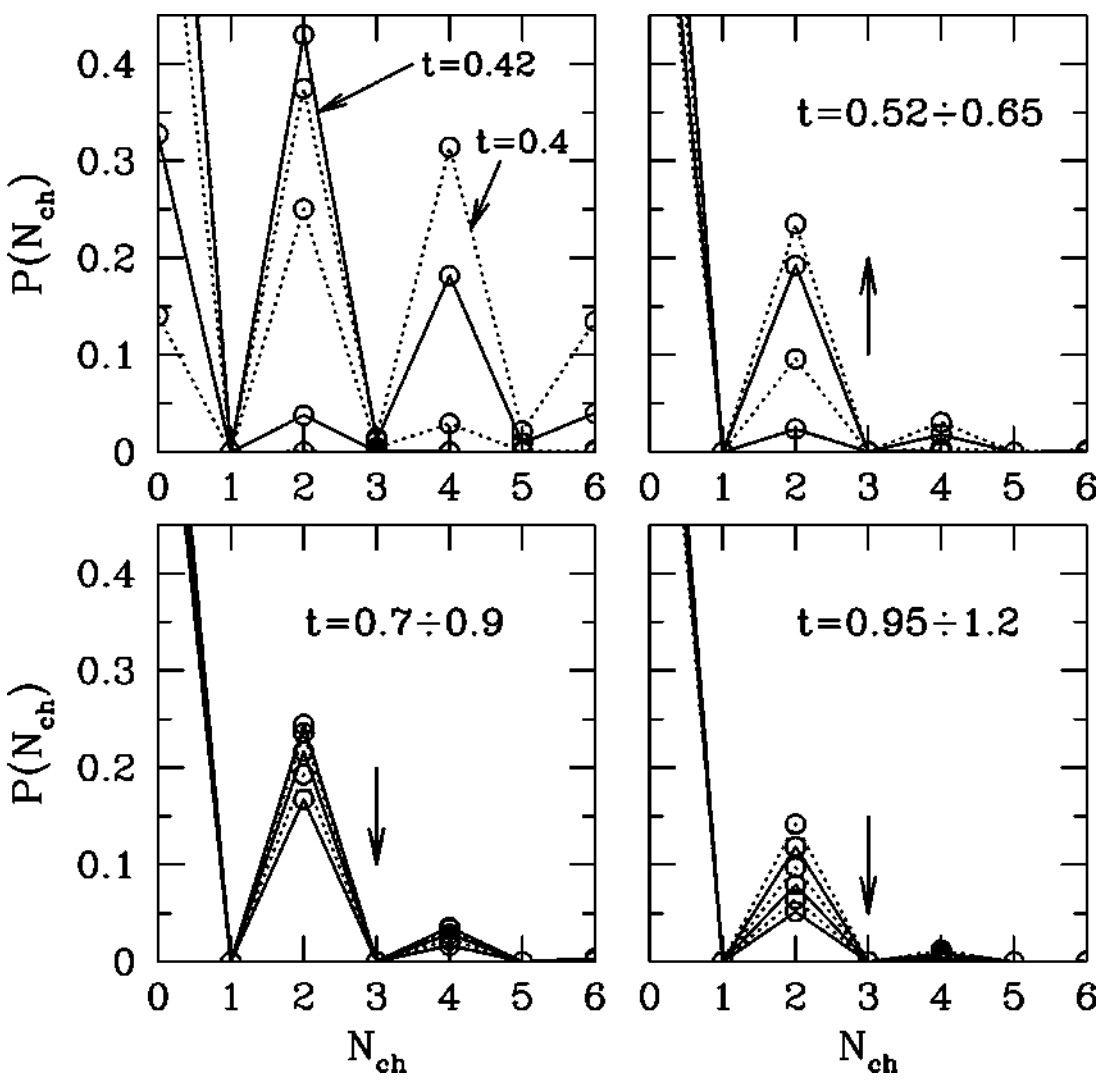

FIG. 10. Truncated fugacity expansion (15) for $\Xi_Q$ (the lattice is $12\times 12$ and $t_{PR}=0.5$): fraction of (micro)states with $N_{ch}$ charges (from 0 to 6). The straight lines between the data points are just drawn for guiding the eye. Left top panel: taken $N_{ch}=2$ as a reference, the plotted lines are relative, from top to bottom, to $t=0.42,0.4,0.45,0.48$, and 0.5. Right top panel: from bottom to top, $t=0.52,0.55,0.6$, and 0.65. Left bottom panel: from top to bottom, $t=0.7,0.75,0.8,0.85$, and 0.9. Right bottom panel: from top to bottom, $t=0.95,1,1.05,1.1,1.15$, and 1.2. Clearly, the probability of observing more than six charges in a $12\times 12$ model system is absolutely negligible when $t>0.45$.

From the knowledge of the pair distribution functions $g_{qq'}$, the charge-charge correlations $\langle q_xq_y\rangle$ are easily determined. Using the identity $q_x=c_{x,1}-c_{x,-1}+2c_{x,2}-2c_{x,-2}$, I find $\langle q_x^2\rangle=2\rho_1+8\rho_2$ and (for $x\neq y$)

$$\begin{aligned}\langle q_xq_y\rangle=&2\rho_1^2[g_{1,1}(x,y)-g_{1,-1}(x,y)]\\ &+8\rho_1\rho_2[g_{1,2}(x,y)-g_{1,-2}(x,y)]\\ &+8\rho_2^2[g_{2,2}(x,y)-g_{2,-2}(x,y)]. \qquad (16)\end{aligned}$$

In deriving the above equation, explicit consideration of charge-inversion symmetry of the Hamiltonian has been made, which implies, also thanks to $\mu_q=\mu_{-q}$, that $\langle c_{x\alpha}c_{y\beta}\rangle=\langle c_{x,-\alpha}c_{y,-\beta}\rangle$.

Charge-inversion symmetry does not hold, in general, for a system with fixed numbers of charges. For instance, $\langle c_{xq}\rangle^{(N_1,N_{-1},N_2,N_{-2})}=N_q/N$ (where the superscript denotes a canonical average). Nevertheless, since the charge configurations that enter the sum (14) are overall neutral, a sum rule such as

$$\sum_{x,y}\langle q_xq_y\rangle^{(N_1,N_{-1},N_2,N_{-2})}=0 \qquad (17)$$

holds for all of the 4-tuples $\{N_q\}$ with $\Sigma_q qN_q=0$. I thus have 13 exact identities of the kind (17) against which the calculation of the canonical $g_{qq'}$ may be checked. I notice, however, that Eq. (16) is useless for computing the canonical charge-charge correlation functions. See, for instance, the case of $\langle q_x q_y \rangle^{(3101)}$. Since $c_{x2}=0$ for all $x$, I find

$$\begin{aligned}\langle q_x q_y \rangle^{(3101)} &= \langle c_{x1} c_{y1} \rangle^{(3101)} - 2\langle c_{x1} c_{y,-1} \rangle^{(3101)} \\ &\quad - 4\langle c_{x1} c_{y,-2} \rangle^{(3101)} + 4\langle c_{x,-1} c_{y,-2} \rangle^{(3101)} \\ &= \langle c_{x1} c_{y1} \rangle^{(3101)} - 2\langle c_{x1} c_{y,-1} \rangle^{(3101)} \\ &\quad - 4\langle c_{x1} c_{y,-2} \rangle^{(3101)} + 4\langle c_{x1} c_{y2} \rangle^{(1310)}, \quad (18)\end{aligned}$$

which is radically different from Eq. (16).

Once the charge-charge correlation functions are known, I can evaluate the static dielectric constant $\epsilon$, which is given, from linear response theory, by [15]

$$\begin{aligned}\epsilon^{-1} &= \lim_{p\to 0} \left\{ 1 - \frac{2\pi}{N T_{\rm Q} p^2} \sum_{x,y} \langle q_x q_y \rangle \exp[-ip(x-y)] \right\} \\ &\simeq 1 - \frac{1}{2\pi T_{\rm Q}} \left\{ N\langle q_0^2 \rangle \right. \\ &\quad \left. + \sum_{z\neq 0} \langle q_0 q_z \rangle \frac{\cos(2\pi z_x/L) + \cos(2\pi z_y/L)}{2} \right\}. \quad (19)\end{aligned}$$

This quantity allows one to distinguish a metal ($\epsilon^{-1}=0$) from an insulator ($\epsilon^{-1}>0$). Coming from low temperature, the vanishing of $\epsilon^{-1}$ (in the thermodynamic limit) will signal a phase transition from an insulating phase, where most of the charges are bound in neutral NN pairs (the medium is polarizable but it is not certainly a conductor) to a metallic phase (where free charge carriers do give rise, in the presence of a driving field, to an electric current). Usually, this transition is driven by the critical unbinding of dipoles, a KT phenomenon in the square-lattice CG model. The fraction of dissociated charges would continuously increase as one moves farther from the transition point into the metallic phase. Within the KT scenario, the inverse dielectric constant actually behaves like an order parameter for the insulator-to-metal transition: in an infinite-sized system, $\epsilon^{-1}$ would jump from $4T_c$ to zero right at the transition temperature $T_c$ (a more-or-less sharp crossover will be observed in a finite system).

An even more direct method for investigating the nature of the DOF phase for the charges is to monitor, as a function of temperature, the average population of various relevant charge arrangements: isolated unit and double charges, pairs of NN sites hosting two opposite charges (1- and 2-dipoles), and neutral trimers (made up of two equal unit charges and one double charge being NN of both). We call $\mathcal{N}_{\rm i1}$ ($\mathcal{N}_{\rm i2}$) the average number of isolated unit (double) charges, $\mathcal{N}_{\rm d1}$ ($\mathcal{N}_{\rm d2}$) the average number of 1-(2-)dipoles, and $\mathcal{N}_{\rm t}$ the average number of neutral trimers. The exact enumeration of these structures is made in parallel to the term-by-term estimate of the partition function. $\mathcal{N}_{\rm d1}$ can be also estimated from the contact value of the pair distribution function $g_{1,-1}$, that is $G_{1,-1}(1)=g_{1,-1}(x,y;|x-y|=1)$, through the relation

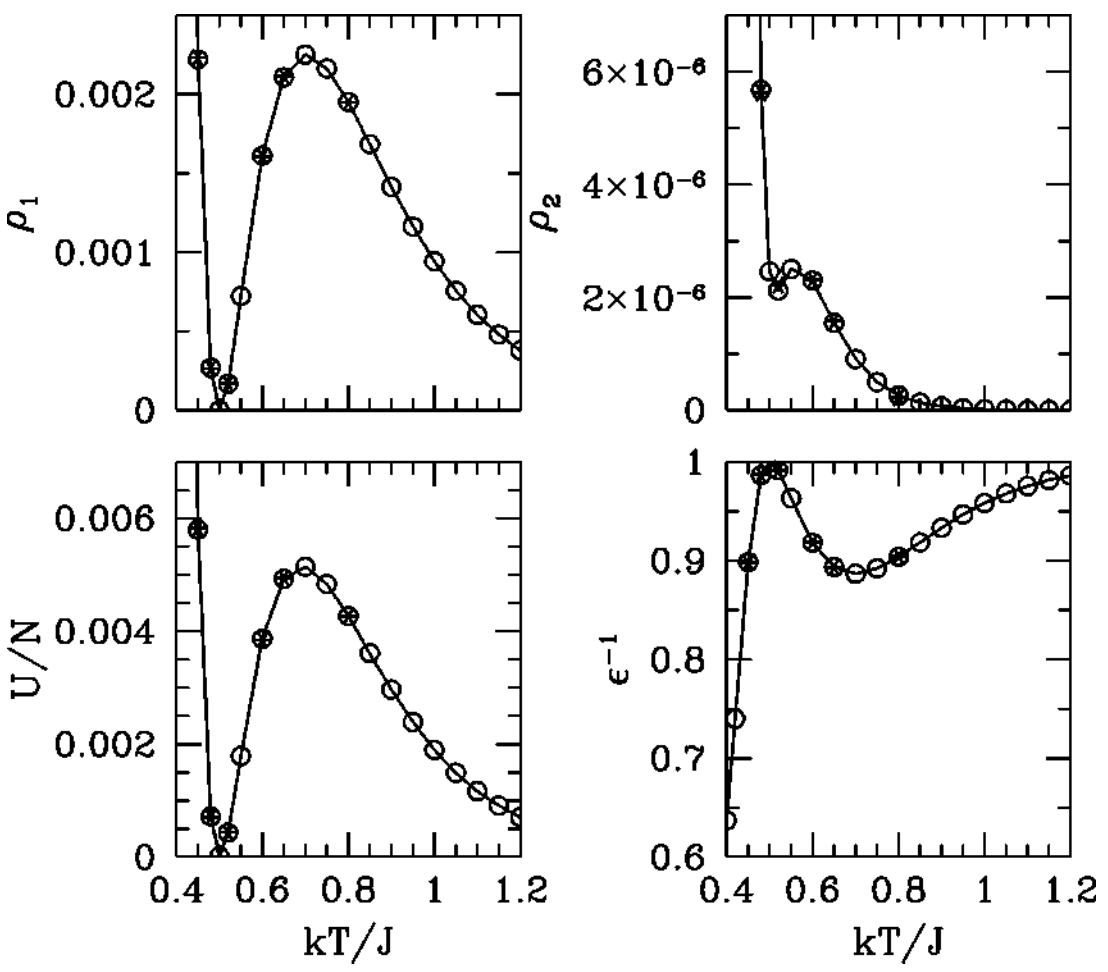

FIG. 11. Thermodynamic properties of the CG model (11)–(13), as it follows from the truncated fugacity expansion (15) of the partition function (the lattice is 12×12 here, with $t_{\rm PR}=0.5$). Exact calculations (○) are compared with MC data (*). Top panels: number densities. Left bottom panel: average energy per site. Right bottom panel: inverse dielectric constant. The density is rather low for both species, particularly for double charges. The nonmonotonic trend of $\epsilon^{-1}$ as a function of $t$ is the symptom of a complex phase behavior with three phases: an insulating phase at high $t$ and two likely metallic phases at low $t$, separated by an insulating state at the PR point.

$$\mathcal{N}_{\rm d1} = \left\langle \sum_{\langle x,y\rangle} (c_{x1}c_{y,-1} + c_{x,-1}c_{y1}) \right\rangle = 4N\rho_1^2 G_{1,-1}(1), \quad (20)$$

and similarly for $\mathcal{N}_{\rm d2}$.

Now, I review the results of the analysis based on Eq. (15) for the 12×12 lattice. The CG parameters corresponding to model A are considered first. I show in Fig. 11 the thermal behavior of the number densities $\rho_1$ and $\rho_2$, of the energy $u$, and of the inverse dielectric constant $\epsilon^{-1}$. Also included for comparison are the outcomes of a MC simulation ($4\times 10^7$ sweeps long). Looking at Fig. 11, what appears first is the strong dilution of the system, which is far more pronounced for double charges than for unit charges. Coming from high $t$, both densities first grow as a result of the increase of $T_{\rm Q}$ (see Fig. 9), until, in the PR region, the rapid drop of $z_1$ causes the vanishing of $\rho_1$ at $t_{\rm PR}$ and the appearance of a local minimum in the profile of $\rho_2$. Past the PR point, the simultaneous increase of $T_{\rm Q}$, $\mu_1$, and $\mu_2$ leads to a rapid filling of the lattice with charges of all types, whence to metallic behavior. Similar to $\rho_1$ is the behavior of $u$, which is throughout positive, denoting the tendency of every charge to keep unlike charges closer than like ones.

Some hints about the nature of the DOF phase for the charges come from the analysis of the behavior of the dielectric constant. When $t$ is high, $\epsilon^{-1}$ is close to 1, indicating insulating behavior. Upon reducing $t$, $\epsilon^{-1}$ gradually lowers until it reaches a minimum at $t=0.7$. Following a recovery at PR, $\epsilon^{-1}$ eventually drops when increasing $T_{\rm Q}$ beyond the PR value. It is too early to say whether these findings would imply metallic behavior for both SOS flat phases (it is nec-essary to wait until a finite-size-scaling study of $\epsilon^{-1}$). Even in this case, however, a curious possibility is that the PR point will actually represent an island of insulating behavior within a sea of metallic behavior (this is consistent with the fact that the surface is rough at PR). Since at PR unit charges are absent, this could be the outcome of a marked tendency of 2 and $-2$ charges to occur in neutral pairs at $t_{\mathrm{PR}}$.

In order to shed some light on the nature of the intermediate phase of the CG model, I present a simplified argument that makes the expectation of two phase transitions rather natural for this system. Suppose that two charges only, 1 and $-1$, are hosted in the lattice (this is a good approximation only at very low $T_{\mathrm{Q}}$). In this case, the contributions to the partition function coming from isolated unit charges and 1-dipoles, respectively, read $W_{\mathrm{i}}=Nz_1^2\Sigma_{|x|>1}\exp(\beta_{\mathrm{Q}}V_{0x})$ and $W_{\mathrm{d}}=4Nz_1^2\exp(\beta_{\mathrm{Q}}V_{01})$. The ratio $R(T_{\mathrm{Q}})=W_{\mathrm{i}}/W_{\mathrm{d}}$ is a monotonously increasing function of $T_{\mathrm{Q}}$ which, irrespective of $z_1$, equals 1 at $t\approx1.0$ (I took $L=24$; however, this result is only weakly dependent on the system size). The same ratio in the event of two opposite double charges only would simply be $R(T_{\mathrm{Q}}/4)$, crossing 1 at one-fourth of the $t$ where $R(T_{\mathrm{Q}})=1$, namely at $t\approx0.25$. Although these numbers are purely indicative, I surmise that, upon decreasing the value of $t$, dissociation of 1-dipoles will come before and separate from the unbinding of 2-dipoles. Should these two events be driving mechanisms of phase transitions, (1) both the intermediate and the high-$T_{\mathrm{Q}}$ phase of the CG model will be metallic (simply because unit charges occur freely in both and thus are able to sustain the electric conduction); (2) the "DOF" metal would be a worse conductor than the "smooth" metal (just because in the DOF phase double charges are frozen in and cannot give rise to a current).

By the way, only a determination of the average number of isolated charges and dipoles that are present in the system can say a definite word about the nature of the three phases of the CG model. My results are reported in Fig. 12 as a function of the SOS temperature $t$. While for both species the amount of isolated charges behaves similarly to the overall density, the number of dipoles shows some differences between unit and double charges. This can be better appreciated by plotting, separately for the two species, the fraction of isolated charges and that of "associated" charges. Since a neutral trimer can either be viewed as a bound pair of 1-dipoles or as a variant of a 2-dipole, I evaluate the total number of bound charges as $\mathcal{N}_{\mathrm{b1}}=2\mathcal{N}_{\mathrm{d1}}+2\mathcal{N}_{\mathrm{t}}$ for unit charges and $\mathcal{N}_{\mathrm{b2}}=2\mathcal{N}_{\mathrm{d2}}+\mathcal{N}_{\mathrm{t}}$ for double charges. I then define (for $\alpha=1,2$)

$$x_{\mathrm{i}\alpha}=\frac{\mathcal{N}_{\mathrm{i}\alpha}}{2N\rho_\alpha}\quad\text{and}\quad x_{\mathrm{d}\alpha}=\frac{\mathcal{N}_{\mathrm{b}\alpha}}{2N\rho_\alpha}. \tag{21}$$

Besides these fractions, it is useful to consider also the fraction $x_{\mathrm{r}\alpha}=1-x_{\mathrm{i}\alpha}-x_{\mathrm{d}\alpha}$ of the residual charges.

Looking at the left top panel of Fig. 13, we observe that $x_{\mathrm{r1}}$ is practically zero for $t>0.5$, meaning that, when $t$ is not too small, isolated and bound charges almost exhaust the total of unit charges. Among the structures that are not monitored, I only mention the NN pairs composed of one double charge and one unit charge of opposite sign (these are similar to isolated unit charges). In fact, $x_{\mathrm{r1}}$ is a bit negative at high $t$, due to an overcounting error in the estimate of $\mathcal{N}_{\mathrm{b1}}$ (if two 1-dipoles have one charge in common, the unit charges are three in total, not four). Coming from high $t$, $x_{\mathrm{i1}}$ and $x_{\mathrm{d1}}$ show a specular trend. While the latter goes down linearly, the former increases until they cross each other at $t\simeq1$. The crossing point would correspond to an insulator-to-metal transition, an event usually referred to as the "unbinding of 1-dipoles."

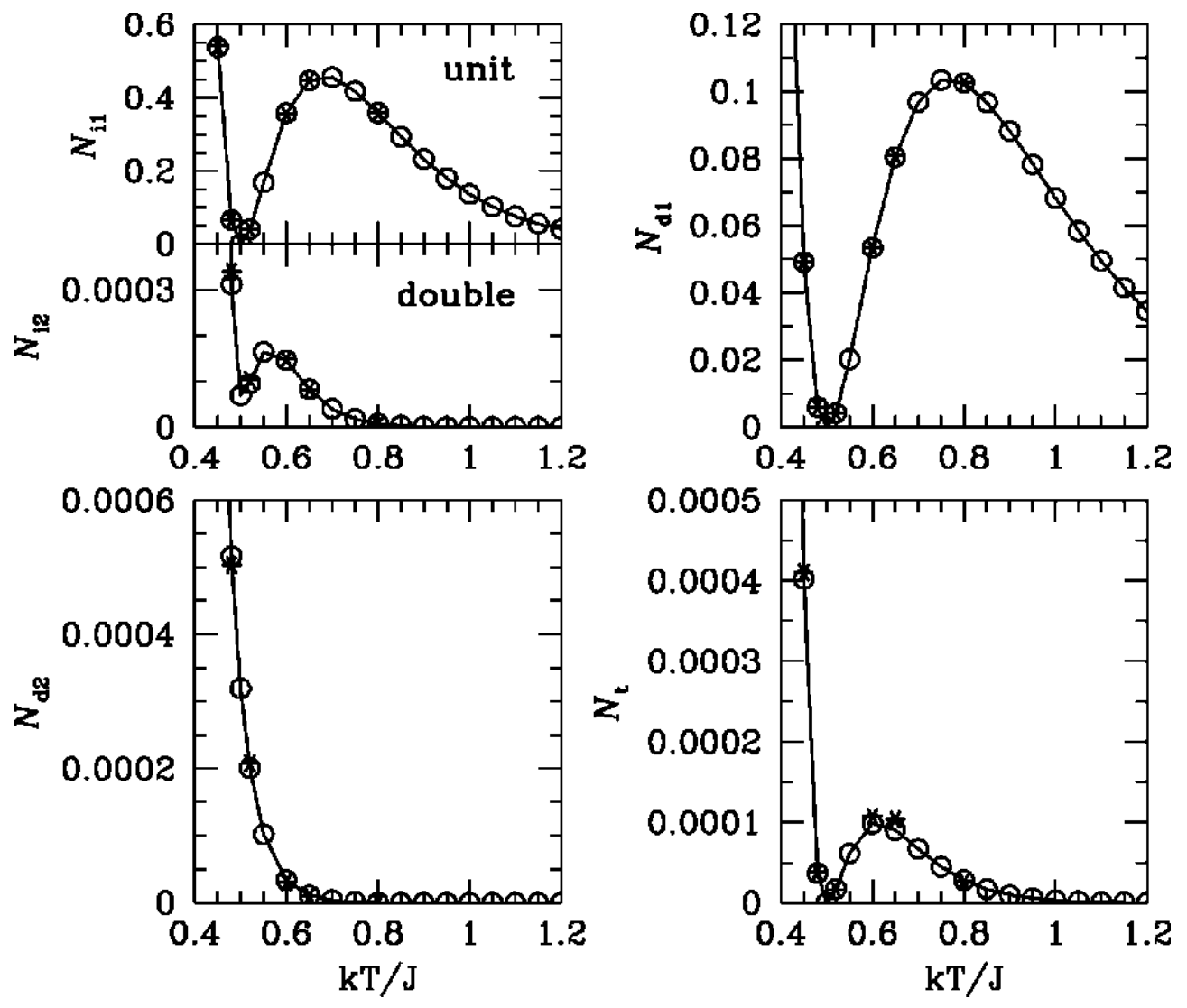

FIG. 12. Statistics of isolated charges, dipoles, and trimers in the CG model (11)–(13), for $L=12$ and $t_{\mathrm{PR}}=0.5$. Exact calculations ($\bigcirc$) from the truncated expansion (15) of the partition function are compared with MC results ($*$) for the same lattice. Left top panel: average numbers $\mathcal{N}_{\mathrm{i1}}$ and $\mathcal{N}_{\mathrm{i2}}$ of the isolated unit (above) and double charges (below). Right top panel: average number $\mathcal{N}_{\mathrm{d1}}$ of 1-dipoles. Left bottom panel: average number $\mathcal{N}_{\mathrm{d2}}$ of 2-dipoles. Right bottom panel: average number $\mathcal{N}_{\mathrm{t}}$ of trimers. I checked that the numbers of 1- and 2-dipoles, as being computed by summing over the 13 4-tuples $\{N_q\}$, are virtually identical to the values drawn from Eq. (20) (and those analogous for 2-dipoles), which uses the contact value of the pair distribution functions.

Moving to double charges, I first note that $x_{\mathrm{r2}}$ is far from being zero (with the only exception of very high $t$ values), indicating that an important category of structures containing double charges was neglected. Evidently, these structures are the above-mentioned $(\pm2,\mp1)$ pairs, as also evidenced by the vanishing of $x_{\mathrm{r2}}$ at $t_{\mathrm{PR}}$. While these structures are irrelevant as for the balance of unit charges, they will contribute a non-negligible fraction to double charges. Anyway, these pairs ought to be included within the class of isolated unit charges (of which they represent a minority), hence they do not enter neither $\mathcal{N}_{\mathrm{i2}}$ nor $\mathcal{N}_{\mathrm{b2}}$.

The overall behavior of $x_{\mathrm{i2}}$ and $x_{\mathrm{d2}}$ is consistent with my expectation that the phase transition at $t_{\mathrm{PR}}$ is promoted by the unbinding of 2-dipoles. For $t>0.6$, $x_{\mathrm{d2}}$ is larger than $x_{\mathrm{i2}}$, but the difference reduces as $t$ goes down. When PR is approached from the DOF phase, the fraction of 2-dipoles obtains further enhancement from the lowering of $z_1$, which in turn causes a reduction in the number of the competing $(\pm2,\mp1)$ bound states. Exactly at PR, the latter are missing and the 2-dipole is a far more preferred configuration for double charges than the isolated state. This explains the insulating character of the CG model right at $t_{\mathrm{PR}}$.

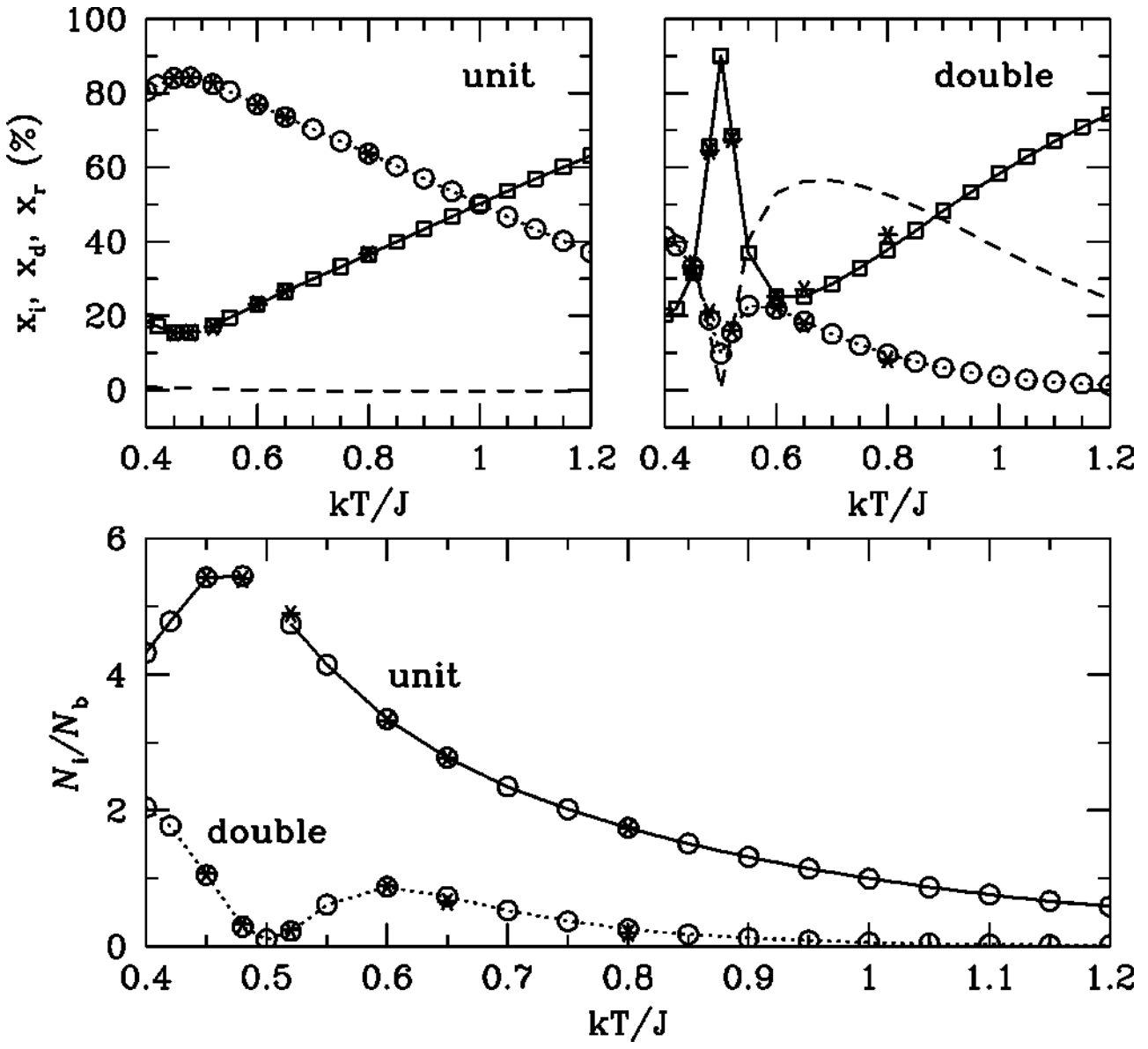

FIG. 13. Statistics of isolated and bound charges in the CG model, as being drawn from the data of Fig. 12. Left top panel: fractions of unit charges (dotted line and ○, $x_{\mathrm{i1}}$; continuous line and □, $x_{\mathrm{d1}}$; dashed line, $x_{\mathrm{r1}}$). Right top panel: same fractions as before but for double charges. Bottom panel: abundancy of isolated charges relative to that of bound charges (continuous line, unit charges; dotted line, double charges). From this picture it appears that dissociation of 1-dipoles occurs at $t \simeq 1$, while most of the double charges remain bound in the DOF phase, up to PR. Hence the DOF phase shows metallic properties, with 2-dipoles and trimers floating in a sea of isolated unit charges.

In the bottom panel of Fig. 13, the charge preference for the isolated state (as opposed to the bound state) is quantified through the plotting of $\mathcal{N}_{\mathrm{i1}}/\mathcal{N}_{\mathrm{b1}}$ and of $\mathcal{N}_{\mathrm{i2}}/\mathcal{N}_{\mathrm{b2}}$. Clearly, the number of free unit charges exceeds that of bound charges for $t < t_{\mathrm{R}}$ and their ratio increases down to PR. Conversely, isolated double charges become as numerous as bound double charges only close to $t = 0.6$, where the DOF character is stronger. However, after reaching a maximum at $t = 0.6$, $x_{i2}/x_{d2}$ goes down again to about zero at PR.

When $t < 0.45$, the results based on Eq. (15) are no longer reliable. However, it is clear from the trend of $\rho_1$ and $\rho_2$ that cluster structures of any kind are being excited now, thus making isolated charges and dipole-like arrangements less and less relevant, as for the overall balance of the charges, than other more complex structures.

Given the above results for the $12\times 12$ system, I conclude that both SOS flat phases are likely metallic in the language of electric charges, owing to the existence of free unit charges in both. In the "DOF" metal, however, double charges exhibit a certain tendency to pairing which is absent in the "smooth" metal. I notice that, more than 10 years ago, a similar conjecture about the CG counterpart of the DOF phase was advanced by Den Nijs [3] who, however, did not provide a numerical demonstration of the kind considered here.

### C. Simulation results

The $12\times 12$ lattice of charges is too small of a system to allow for neat phase-transition signatures. On the other hand, when considering much larger systems, a low-fugacity expansion of the partition function is no longer a viable solution and a different strategy is in order. In these cases, MC simulation is the only available method. However, the very same feature that makes it possible to perform the perturbative analysis (that is, a strong dilution of the charges) is also the weak point of MC sampling: very long runs must be carried out in order to collect sufficient statistics.

There are at least two ways to implement the MC method in the present CG model. One solution is similar to that described in Ref. [15]. A single MC step is articulated as follows: first, I randomly choose a pair of NN or NNN lattice sites. Then, the charge at one site is increased by one or two, whereas the charge at the other site is decreased by the same amount. Should by this means one obtain a charge different from $0,\pm 1$, or $\pm 2$, the move will be rejected (like it would be if an enormously positive chemical potential were associated with this charge). Otherwise, the energy change is calculated and the move is accepted or rejected according to the usual Metropolis rule.

An alternative to the above algorithm is the following: first, a pair of NN or NNN sites is chosen at random and their charge contents are kept; then, depending on the values of these charges, a MC move out of the following list is attempted (note that the total charge is conserved anyway): (1) a charge is moved to an empty site; (2) one double charge is broken into two unit charges; (3) two equal unit charges merge into one single double charge; (4) two opposite charges are created; (5) two opposite charges are destroyed; and (6) the positions of one double charge and of one unit charge of opposite sign are interchanged. Next, the trial move is accepted or rejected depending on the Metropolis weight. I point out that the above listed elementary moves are characterized by different *a priori* probabilities, which then obliges one to modify the usual form of the Metropolis acceptance probability.

We have checked by intensive MC runs for a $12\times 12$ system that the two algorithms above give indeed practically the same results, which are also fully consistent with the exact calculations (see Figs. 11–13). Moreover, the performance of both algorithms is similar, whereas the acceptance of MC moves is about twice as large for the second of the two.

For the simulation, three system sizes are considered, $L = 48,72$, and 96. The same parameters as in Eq. (5) (model A) are used first. After equilibration, as many as $2\times 10^6$ sweeps are generated, each sweep consisting of one MC step per site (in the high-$t$ region, a longer MC trajectory of $4\times 10^6$ sweeps is generated for $L = 48$ and 96 in order to achieve better statistics for the double charges). Averages are updated every 10 sweeps. Among the quantities that can help to understand the way how charges are distributed on the lattice, the following are especially monitored: the number densities $\rho_q$, the average energy per site $u$, the radial distribution functions $G_{qq'}(r)$, and the inverse dielectric constant $\epsilon^{-1}$. During the run, I also compute the statistics of isolated

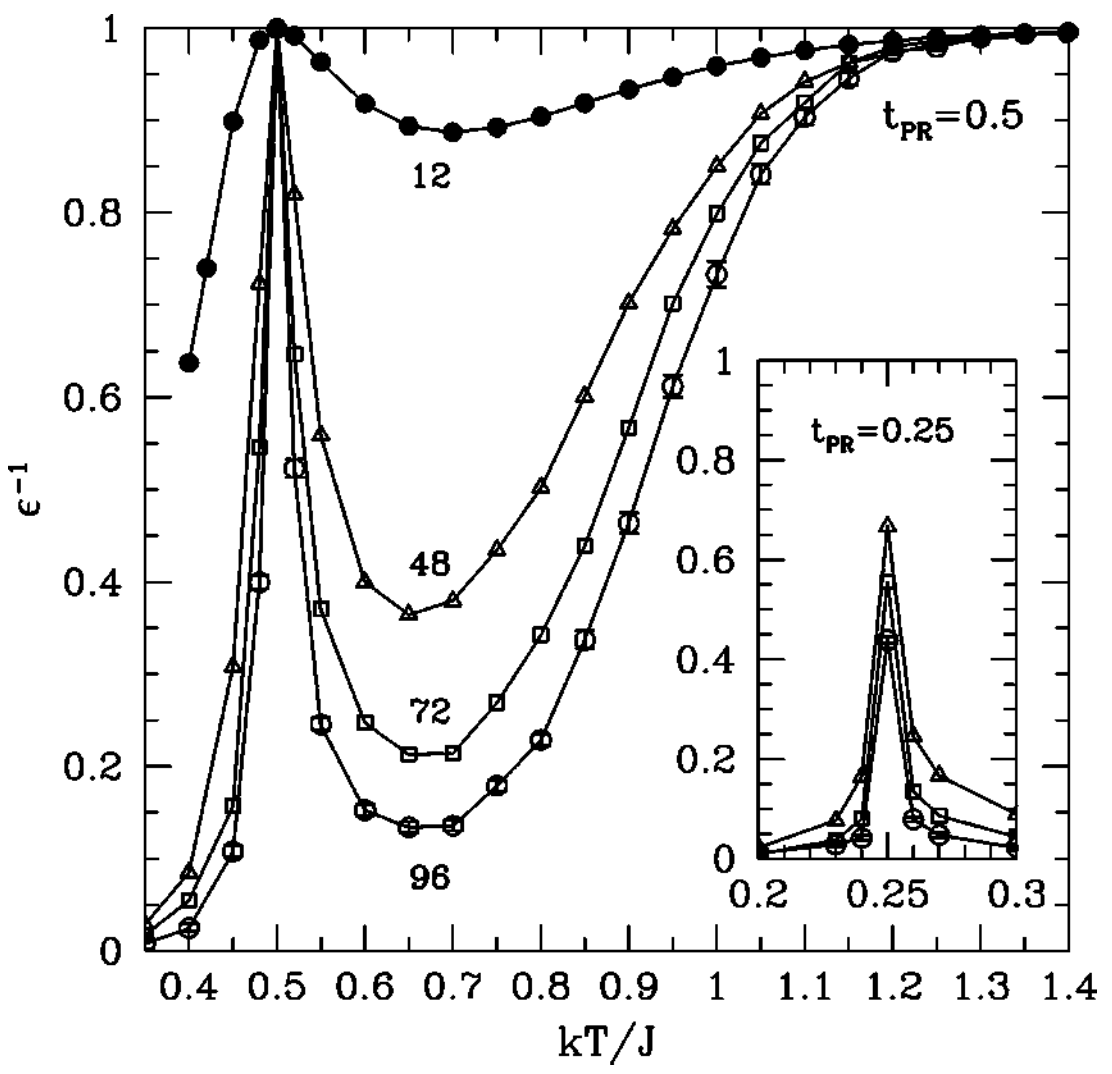

FIG. 14. Inverse dielectric constant $\epsilon^{-1}$ in the CG model (11)–(13): a comparison is made between $t_{\rm PR}=0.5$ (main picture) and $t_{\rm PR}=0.25$ (inset). Data are shown for a number of $L$ values: 12 (full circles, exact calculation—same data as in Fig. 11), 48 (△), 72 (□), and 96 (○). The scaling behavior of $\epsilon^{-1}$ clearly indicates that the system is always metallic when the dual SOS surface is flat. However, when $t_{\rm PR}=0.5$, it is insulating at PR, as well as in the whole rough phase. Conversely, when $t_{\rm PR}=0.25$, the dielectric character is probably metallic also at PR.

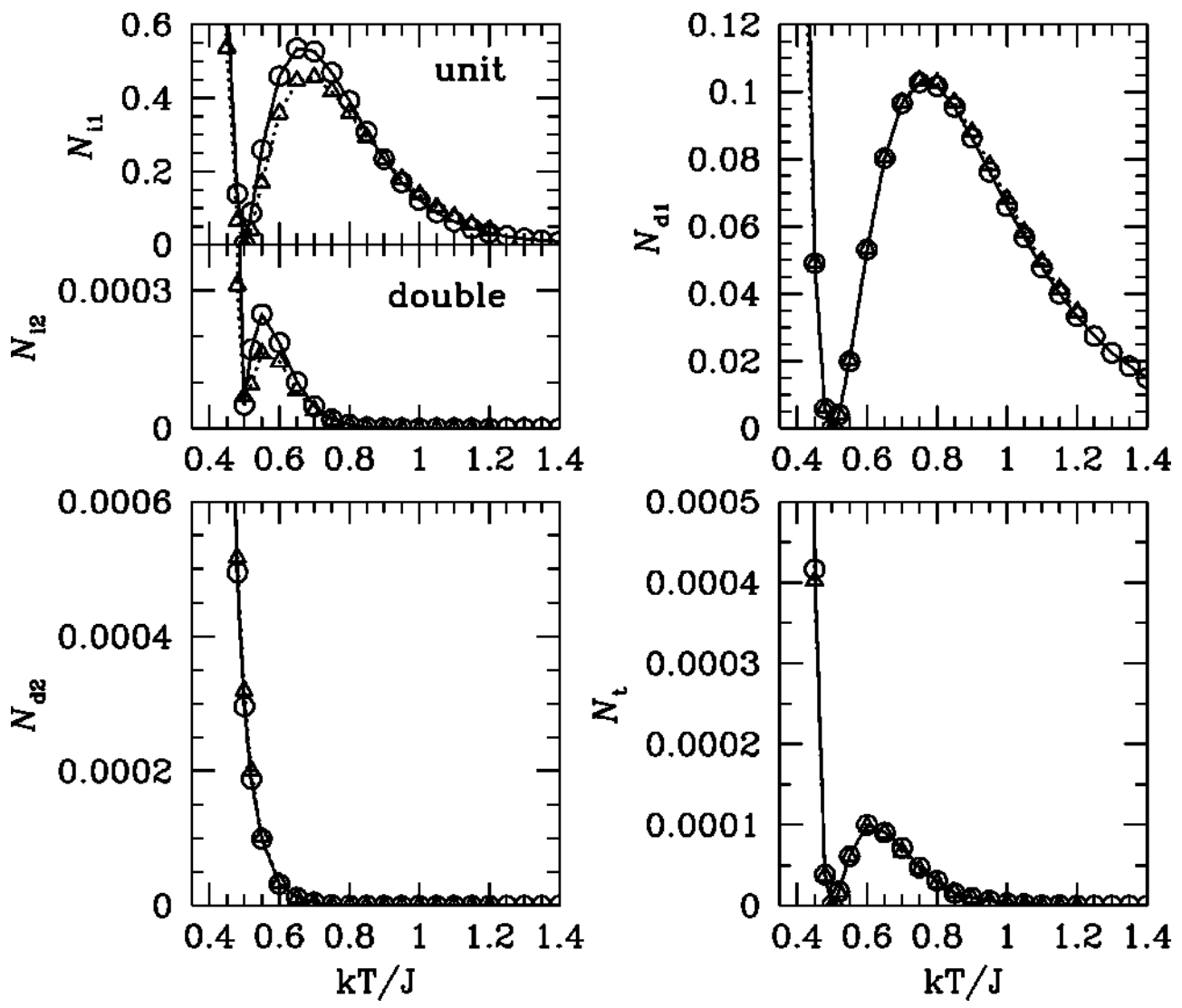

FIG. 15. Statistics of isolated charges, dipoles, and trimers in the CG model (11)–(13), for $t_{\rm PR}=0.5$. The exact calculations for $L=12$ (△) are compared with MC simulation results for $L=96$ (○) (to allow for a better comparison, the data for $L=96$ have been divided by 64). Left top panel: average numbers of isolated unit (above) and double charges (below). Right top panel: average number of $(1,-1)$ NN pairs. Left bottom panel: average number of $(2,-2)$ NN pairs. Right bottom panel: average number of $(\pm 2,\mp 1,\mp 1)$ trimers (where the two unit charges are both NN of the double charge).

charges, of 1- and 2-dipoles, and of neutral trimers.

In Fig. 14, the inverse dielectric constant is plotted for all sizes. It is rather evident where the trend goes when increasing the system size: $\epsilon^{-1}$ eventually vanishes in both SOS flat phases but not at PR. Therefore, as already anticipated, the CG system is metallic both in the smooth and in the DOF phase (it remains to be seen what distinguishes between the "DOF" and the "smooth" metal). Surprisingly, however, the same system is insulating at the (isolated) PR point between the two. When the PR transition is first-order (model B), the nonzero maximum of $\epsilon^{-1}$ at $t_{\rm PR}=0.25$ appears to be only a finite-size effect (Fig. 14, inset). This suggests that the system is metallic also at $t_{\rm PR}$.

Going back to model A, the metal-insulator transition at $t_{\rm R}$ can be located through the criterion $\epsilon^{-1}(T_c)=4T_c$. This gives $t_{\rm R}\simeq 1.3$, which is consistent with the overall behavior of $\epsilon^{-1}$ for model A, not as much with the MC data of Fig. 1 (unless I admit that $L=72$ is still too small a size).

In Fig. 15 I compare the $L=12$ and $L=96$ systems as far as the average numbers of isolated charges, of dipoles, and of neutral trimers are concerned. To make this comparison more significant, the numbers for $L=96$ are divided by 64, which is the ratio between the two respective $N$ values. What immediately stands out is the similarity of behavior between the two sizes, the main difference lying in the statistics of isolated charges, which are comparatively more numerous in the bigger of the two lattices.

The overall behavior of $x_{i\alpha}$ and $x_{d\alpha}$ for $L=96$ is the same as for $L=12$ (see the top panels of Fig. 16; observe that, beyond $t=1.15$, no double charge appeared during the run). In the same figure, I compare the relative abundancy of isolated and bound charges for $L=12$,48, and 96. Amazingly, no really new feature shows up in the behavior of the largest sizes that is not already present in the $L=12$ system. Notwithstanding for $t>0.7$ the accuracy of my MC estimate of the extremely small density of double charges is very poor, it is clear that the infinite-size behavior of the CG model is already well accounted for by the tiny $12\times 12$ lattice.

Upon decreasing $t$ beyond the roughening value, the average number of isolated double charges smoothly grows, with respect to the number of associated charges, until a maximum relative abundancy of about 1 is attained for $t=0.6$. As I move towards the PR point, however, the insulating character reappears due to a drop in the number of the $(\pm 2,\mp 1)$ pairs. Eventually, all kinds of excitations become permitted in the smooth, fully metallic phase, not just isolated charges, which explains the trend observed for $t<0.5$. Hence I confirm that, due to a large fraction of free unit charges, the DOF phase has a metallic counterpart in terms of charges, likewise the smooth phase. However, in the "DOF" metal a large portion of double charges are bound, which is not the case for the "smooth" metal.

As far as model B is concerned, I report in Fig. 17 the statistics of isolated and bound charges in the $L=48$ and $L=96$ lattices (same analysis as illustrated in Fig. 16 for model A). The main difference from model A lies in the behavior of double charges near PR. The fraction of bound charges never grows beyond 50%, even at PR. As a result, the charged mixture is metallic also at PR. The first-order character of the PR transition is revealed by the scarce size

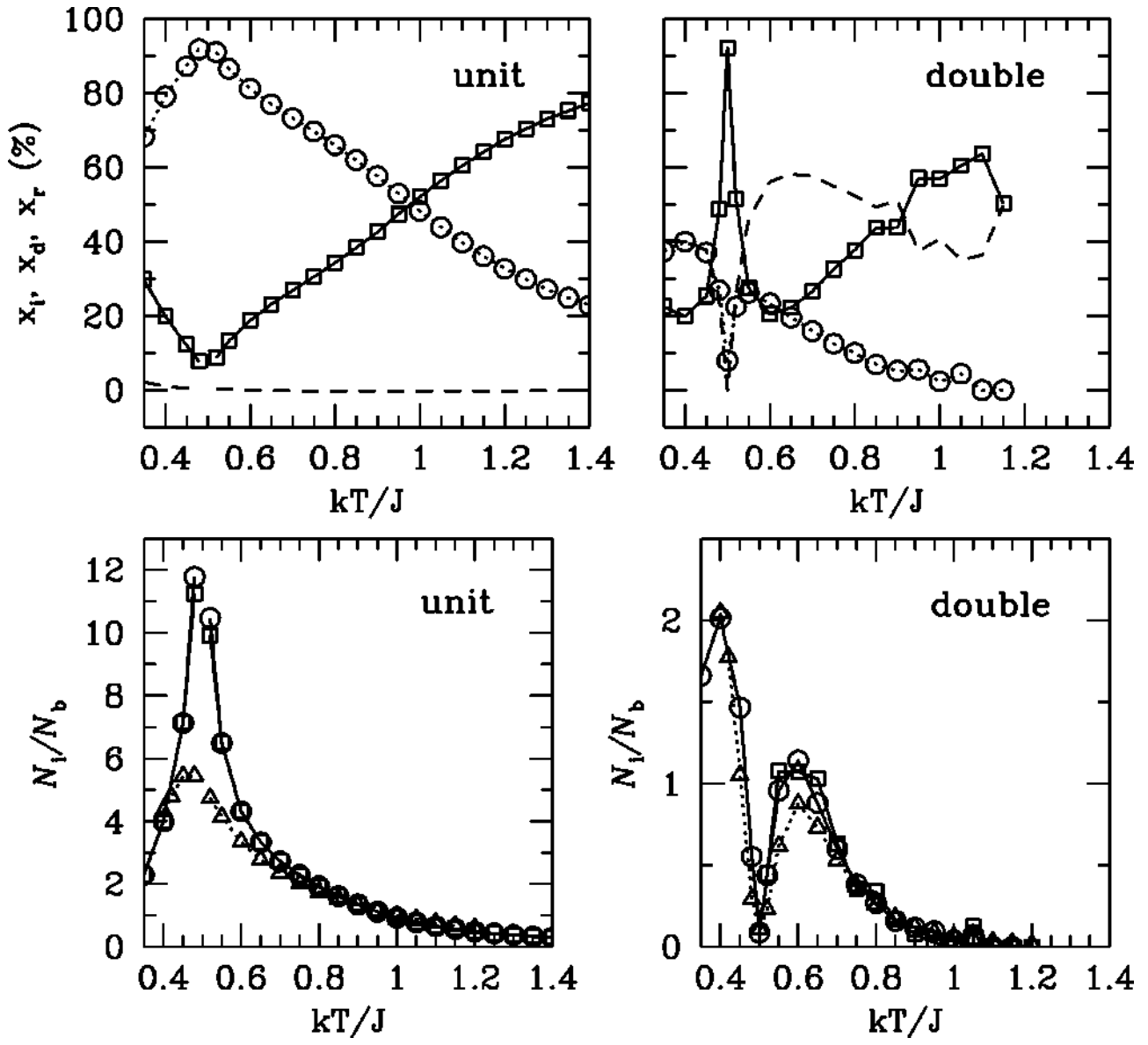

FIG. 16. Statistics of isolated and bound charges in the CG model that is dual to the SOS model A, for $L=96$. Left top panel: fractions of unit charges (dotted line and ○, $x_{i1}$; continuous line and □, $x_{d1}$; dashed line, $x_{r1}$). Right top panel: same fractions as before but for double charges. Bottom panels: three system sizes, $L=12$ (dotted line and △), 48 (□), and 96 (○), are compared as for the abundancy of isolated vs bound charges (left, unit charges; right, double charges). The conclusions drawn for the smallest size are substantially confirmed in the larger systems: while dissociation of 1-dipoles occurs nearby $t=1$, 2-dipoles do not unbind until the PR point is reached.

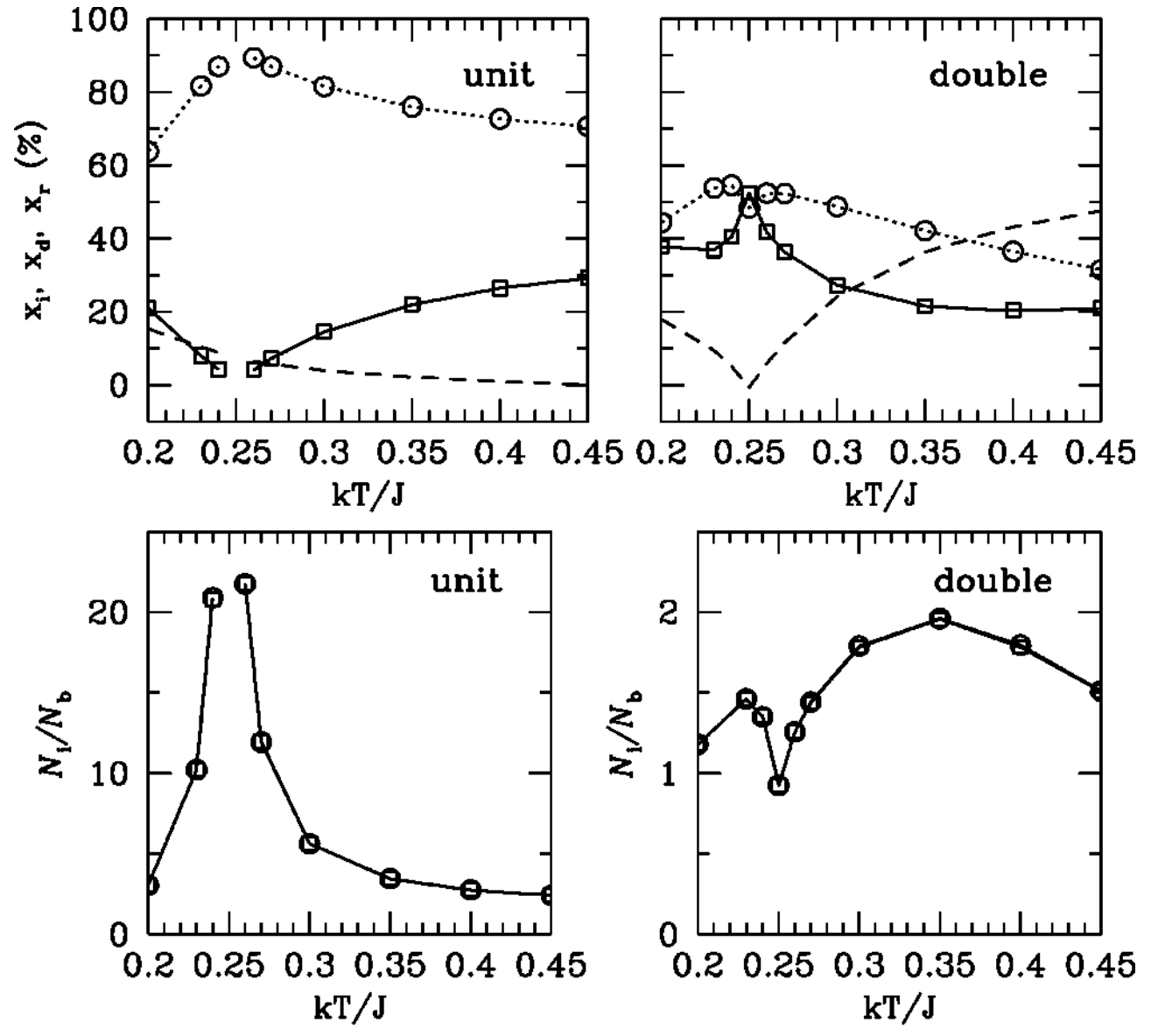

FIG. 17. Statistics of isolated and bound charges in the CG model, for $t_{\mathrm{PR}}=0.25$ (model B) and $L=96$. Left top panel: fractions of unit charges (dotted line and ○, $x_{i1}$; continuous line and □, $x_{d1}$; dashed line, $x_{r1}$). Right top panel: same fractions as before but for double charges. Bottom panels: two lattice sizes, $L=48$ (□) and $L=96$ (○), are compared as for the relative abundancy of isolated and bound charges (left, unit charges; right, double charges). Unlike the previous case (model A), a non-negligible abundancy of isolated double charges at PR causes the mixture to be metallic also at PR.

dependence of the charge abundancies, as shown in the bottom panel of Fig. 17.

### D. Entropy of the charges

It is by now natural to analyze the phase transitions undertaken by our fluid of interacting charges in terms of the so-called residual multiparticle entropy (RMPE). Since the publication of [16], a lot of calculations have shown [17] the intrinsic validity for many model systems of a criterion, hereafter referred to as the entropic criterion, aimed at inferring the amount of configurational order that is present in a fluid system from the importance of many-body spatial correlations in the overall entropic balance. After the lattice implementation of the formula for the RMPE [18], the entropic criterion has proved to be useful also for lattice systems [19].

I recall that the RMPE is the difference between the total system entropy and its lowest-order contributions in the grand-canonical many-particle correlation expansion [18], namely the ideal-gas and the two-body terms which, for a mixture of lattice particles, respectively, read as

$$\frac{S^{(0)}}{k}=\ln(1+2z_1+2z_2)-2\rho_1\ln z_1-2\rho_2\ln z_2 \qquad (22)$$

and

$$\frac{S^{(2)}}{k}=-\sum_{q,q'}\rho_q\rho_{q'}\sum_{x<y}[g_{qq'}(x,y)\ln g_{qq'}(x,y)-g_{qq'}(x,y)+1]. \qquad (23)$$

Note that $S^{(0)}$ is the entropy of an ideal-gas mixture of four species with activities $z_1, z_1, z_2$, and $z_2$ and no constraint on the particle numbers relative to each other. Moreover, the values of $z_1$ and $z_2$ are such as to reproduce in the ideal-gas mixture the same densities as for the interacting system [18], a prescription leading to

$$z_1=\frac{\rho_1}{1-2\rho_1-2\rho_2} \quad \text{and} \quad z_2=\frac{\rho_2}{1-2\rho_1-2\rho_2}, \qquad (24)$$

where $2\rho_1$ and $2\rho_2$ are the overall densities of unit and double charges, respectively. The expression (24) for $S^{(2)}$ is a sum, over all pairs of charge species and all ordered pairs of lattice sites, of the same two-body term appearing in the entropy expansion for the continuum.

While the calculation of $S^{(2)}$ only requires the knowledge of the pair distribution functions, the total entropy $S$ of the charges cannot be directly obtained in a MC experiment. However, it can be evaluated anyway by the method of thermodynamic integration, which gives $S$ in terms of the energy and the number densities of the charges along a path in $T_{\mathrm{Q}}$.

In order to derive a formula for the entropy as a function of the SOS temperature, I start from the grand-canonical Massieu function $\tilde{S}$, given by

$$\frac{\tilde{S}}{k}=\frac{S}{k}-\beta_{\rm Q}U+\beta_{\rm Q}\mu_1 N_1+\beta_{\rm Q}\mu_2 N_2. \qquad (25)$$

Since my MC data were collected for a path in $T_{\rm Q}$ at fixed $C, t_{\rm PR}$, and $y_4$, I need to calculate the total derivative of $\tilde{S}$ with respect to $\beta_{\rm Q}=1/T_{\rm Q}$, where $\mu_1$ and $\mu_2$ are viewed as functions of $\beta_{\rm Q}$. Using the chain rule of derivation and standard thermodynamic relations, I readily obtain

$$\frac{d(\tilde{S}/k)}{d\beta_{\rm Q}}=-U+N_1\left(\mu_1+\beta_{\rm Q}\frac{d\mu_1}{d\beta_{\rm Q}}\right)+N_2\left(\mu_2+\beta_{\rm Q}\frac{d\mu_2}{d\beta_{\rm Q}}\right). \qquad (26)$$

Considering that

$$\mu_1(\beta_{\rm Q})=\frac{1}{\beta_{\rm Q}}\ln\left(\frac{\pi C}{2k}\left|\frac{t_{\rm PR}}{\beta_{\rm Q}}-\frac{1}{\pi}\right|\right) \qquad (27)$$

and

$$\mu_2(\beta_{\rm Q})=\frac{1}{\beta_{\rm Q}}\ln\left(\frac{\pi y_4}{2\beta_{\rm Q}J}\right), \qquad (28)$$

the derivatives of $\mu_1$ and $\mu_2$ will read as

$$\frac{d\mu_1}{d\beta_{\rm Q}}=-\frac{\mu_1}{\beta_{\rm Q}}-\frac{t_{\rm PR}}{\beta_{\rm Q}^2 t_{\rm PR}-\beta_{\rm Q}^3/\pi} \qquad (29)$$

and

$$\frac{d\mu_2}{d\beta_{\rm Q}}=-\frac{1+\beta_{\rm Q}\mu_2}{\beta_{\rm Q}^2}. \qquad (30)$$

Upon inserting Eqs. (29) and (30) into Eq. (26), I thus get

$$\frac{d(\tilde{S}/k)}{d\beta_{\rm Q}}=-U-\frac{\pi t_{\rm PR}T_{\rm Q}^2}{\pi t_{\rm PR}T_{\rm Q}-1}N_1-T_{\rm Q}N_2. \qquad (31)$$

The subsequent use of Eq. (25) finally yields the following expression for the entropy:

$$\frac{S(T_{\rm Q})}{k}=\frac{1}{T_{\rm Q}}(U-\mu_1N_1-\mu_2N_2)+\int_0^{T_{\rm Q}}\left[U(T)+\frac{\pi t_{\rm PR}T^2}{\pi t_{\rm PR}T-1}N_1+TN_2\right]\frac{dT}{T^2}. \qquad (32)$$

The calculated entropies are shown in Fig. 18 for the CG parameters corresponding to model A. Two distinct sizes are compared here, $L=12$ and $L=96$. In the $L=12$ case, all plotted quantities are the outcome of an analytic calculation, which is made possible by the exact knowledge of the partition function and of the pair distribution functions. Anyway, I have verified that the independent $S$ estimate through Eq. (32) gives exactly the same result (apart from an undetermined constant, see below). In the other case, $L=96$, the entropy and its lowest-order terms in the multiple-correlation expansion are numerically computed via Eq. (32). This notwithstanding, my MC sampling is so accurate that the minute difference giving the RMPE is an extremely smooth function of the SOS temperature (see Fig. 18), even close to the highest temperature ($t=1.15$) where I am able to quantify the number of double charges during the simulation and, therefore, to compute $S^{(0)}$. I point out that the total entropy of the $96\times96$ system would in principle convey an undetermined overall constant, say its value at $t=1.4$, which I have arbitrarily set equal to $S(1.4)$ of the $12\times12$ lattice. In turn, this indeterminacy is transferred to the RMPE which, for the largest lattice, is only fixed up to an unknown (but likely very small) additive constant.

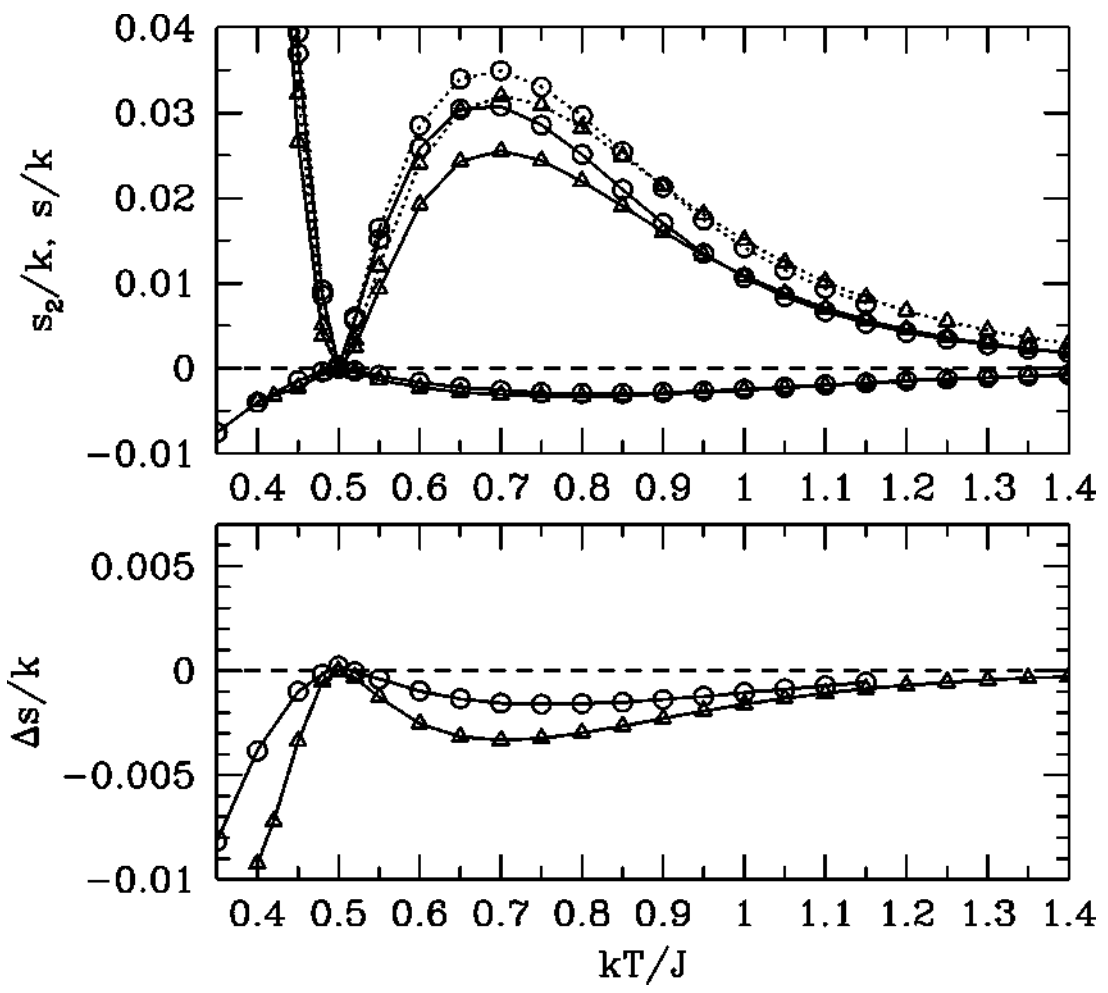

FIG. 18. Residual multiparticle entropy (RMPE) of the CG model (11)–(13), for $t_{\rm PR}=0.5$ (model A). Two sizes are compared, $L=12$ (△, exact calculation) and $L=96$ (○, MC simulation). In the above picture, I show three quantities for each size: total entropy per site (the continuous lines above the $x$ axis), ideal-gas entropy per site (the dotted lines), and two-body entropy per site (the continuous lines below the $x$ axis). Below: RMPE for $L=12$ (△) and 96 (○). The RMPE shows the expected behavior: it takes negative values in both (fluid) metallic phases, apparently vanishing at PR. It eventually moves towards zero when approaching the roughening point. Looking at the picture below, I can hardly say whether the RMPE of an infinite-sized system would be indeed positive in the insulating/rough regime.

The behavior of all the entropies is nonmonotonic in $t$, following somehow the thermal evolution of the densities. Coming from low $T_{\rm Q}$, the multiplicity of the system macrostate first grows. On approaching PR, the dip of $z_1$ becomes evident, leading to an abrupt fall in the number of microstates, hence in the entropy. Upon going across the PR point, the entropy grows again, but now much more rapidly than before. The absolute value of $S^{(2)}$ shows a similar behavior.

Now, I attempt an interpretation of the RMPE profile. Usually, the RMPE of a simple fluid that undergoes a unique ordering phase transition (induced by varying either the temperature at fixed density or the density at fixed temperature) is found to be negative in the disordered phase (low density/

high temperature), while becoming eventually positive on approaching the ordered phase. In fact, the RMPE is generally found to vanish very close to the transition point (with a tolerance of a few percent), in such a way that the location of the RMPE zero is a rather good estimate, sometimes a very good one, of the transition point.

Considering the results of Fig. 18, I conclude that, even for a charged fluid mixture on a lattice, the behavior of the RMPE as a function of $t$ is highly informative of the general structure of the phase diagram. In fact, the RMPE distinctly vanishes at $t_{\rm PR}$ and smoothly approaches zero from below near $t_{\rm R}$; moreover, it is throughout negative below $t_{\rm R}$, in agreement with the fact that, in both metallic phases, the charges are completely (for $t<t_{\rm PR}$) or at least partially disordered (for $t>t_{\rm PR}$). In particular, in the DOF regime where the double charges show a preference towards dipole-like pairing, a minor degree of disorder (as compared with the full-metal regime) goes along with a less negative RMPE value. As a matter of fact, I cannot say whether the apparent failure of the entropic criterion in accounting for the KT metal-insulator transition (the RMPE remains negative beyond $t_{\rm R}$) is a real drawback of the criterion or is in fact just a finite-size effect, also complicated by the unknown constant value I was alluding to before and by my inability to count double charges beyond $t\simeq 1.15$.

## IV. CONCLUSIONS

In this paper, the thermodynamics of a SOS surface model undergoing, besides the usual roughening, a PR phase transition is exactly mapped onto the grand-canonical ensemble of a 2D lattice CG model of unit and double charges. Upon adjusting the SOS model parameters, it is possible to make PR (as well as the analogous transition in the CG model) first-order. Both models have been studied mainly through MC simulation, supplemented in the CG case by exact finite-size calculations.

The duality between the two models has actually served as the expedient to investigate the possibility of a more complex phase behavior than usual in a gas of lattice charges, generally exhibiting a unique phase transition from an insulator at low temperature to a metal at high temperature. In particular, I have been able to describe a new kind of phase transition between two different metallic phases which are the counterpart of the smooth and the DOF surface phases. Precisely, while the cold, fragile "DOF" metal is one where the unit charges are free and a large fraction of the double charges are bound, all kinds of excitations (not just isolated charges) become permitted in the hot, strong "smooth" metal. Hence the PR transition becomes translated, in the CG language, into the unbinding of dipoles being formed by double charges. PR itself, when second-order, is an isolated point in an otherwise metallic regime where the charged mixture displays insulating character, such as we find only at low temperature.

## ACKNOWLEDGMENTS

Helpful discussions with Paolo V. Giaquinta and Giancarlo Trimarchi are gratefully acknowledged.

## APPENDIX: MAPPING OF THE SOS MODEL ONTO A LATTICE COULOMB GAS

In this Appendix I provide the proof that the SOS model at Eq. (3) and the CG model defined by Eqs. (11)–(13) are dual to each other. By that I mean that the two partition functions are the same, up to an unimportant multiplicative constant, even though with inverted temperature scales.

First, I denote by $hWh\equiv\Sigma_{x,y}h_xW_{x,y}h_y$ the purely Gaussian part in Eq. (3) as multiplied by $\beta\equiv 1/(kT)$. The kernel $W_{x,y}$ precisely reads as

$$W_{x,y}=\beta J[(4+\kappa^2)\delta_{x,y}-\delta_{|x-y|,1}]. \tag{A1}$$

In the above formula, a regularization parameter $\kappa^2$ has been introduced in order to make all eigenvalues of $W$ strictly positive (this factor will be later sent to zero so as to eventually recover the original model).

Given Eq. (A1), the SOS partition function is given by

$$Z_{\rm sG}=\lim_{\kappa^2\to 0}\int \mathcal{D}h\,\exp(-hWh)\times\prod_x\left[\frac{1+\beta y_2\cos(2\pi h_x)+\beta y_4\cos(4\pi h_x)}{1+\beta y_2+\beta y_4}\right]. \tag{A2}$$

Each factor in the above product can be rearranged as follows:

$$\frac{1+\beta y_2\cos(2\pi h_x)+\beta y_4\cos(4\pi h_x)}{1+\beta y_2+\beta y_4}=\sum_{q_x=0,\pm1,\pm2} z_{q_x}\exp(2\pi i q_x h_x), \tag{A3}$$

where

$$z_{q_x}=(1+\beta y_2+\beta y_4)^{-1}\left(\frac{y_2}{|y_2|}\right)^{q_x}\exp(\beta_{\rm Q}\tilde{\mu}_{q_x}q_x^2), \tag{A4}$$

being

$$\beta_{\rm Q}=\frac{\pi}{\beta J},\ \exp(\beta_{\rm Q}\tilde{\mu}_{\pm1})=\frac{\beta|y_2|}{2},\qquad \exp(4\beta_{\rm Q}\tilde{\mu}_{\pm2})=\frac{\beta y_4}{2}. \tag{A5}$$

In Eq. (A3), the sum indexes $q_x$ are later interpreted as integer electric charges. Using the above representation, the product at Eq. (A2) can be reshuffled to read finally as

$$(1+\beta y_2+\beta y_4)^{-N}\sum_{\{q_x\}}\left(\frac{y_2}{|y_2|}\right)^{\Sigma_x q_x}\exp\left(\beta_{\rm Q}\sum_x\tilde{\mu}_{q_x}q_x^2\right)\times\exp\left(2\pi i\sum_x q_xh_x\right). \tag{A6}$$

After an obvious Gaussian integration, the $\kappa^2 \rightarrow 0$ limit is taken, with the result that the only nonzero contributions to the SOS partition function come from the $\{q_x\}$ configurations satisfying the condition $\Sigma_x q_x = 0$. The final outcome is

$$Z_{\rm sG} \propto \sum_{\{q_x\}}{}' \exp\left(\beta_{\rm Q} \sum_x \tilde{\mu}_{q_x} q_x^2\right) \exp\left(-\beta_{\rm Q} \sum_{x<y} V_{x,y} q_x q_y\right) \equiv \Xi_{\rm Q}, \tag{A7}$$

where

$$V_{x,y} = \frac{\pi}{N} \sum_{p \neq 0} \frac{\exp[ip(x-y)] - 1}{2 - \cos p_x - \cos p_y} \tag{A8}$$

is just the square-lattice Coulomb potential ($p$ is a Born-Von Karman vector). The ‘‘prime’’ over the sum in Eq. (A7) is there to recall that only the neutral charge configurations are included into the sum.

The right-hand side of Eq. (A7) is the grand-canonical partition function of an overall neutral system of 2D lattice charges (obviously, $q_x = 0$ means that no charge is present at the $x$ site). Hence this system shows the same number of phases and phase transitions as in the original SOS model. Note that the chemical potential is $\mu_1 = \tilde{\mu}_1$ for $\pm 1$ (unit) charges and $\mu_2 = 4\tilde{\mu}_2$ for $\pm 2$ (double) charges. There are no charges of magnitude greater than two.

I finally point out that, notwithstanding that the original model is badly defined at low temperatures (see Sec. II A), the model defined by Eqs. (A7) and (A8) has a proper definition whatever the value of $\beta_{\rm Q}$.

[1] S.T. Chui and J.D. Weeks, Phys. Rev. B **14**, 4978 (1976).
[2] M. Den Nijs and K. Rommelse, Phys. Rev. B **40**, 4709 (1989).
[3] M. Den Nijs, Phys. Rev. Lett. **64**, 435 (1990).
[4] H.S. Youn and G.B. Hess, Phys. Rev. Lett. **64**, 918 (1990); P. Day, M. LaMadrid, M. Lysek, and D. Goodstein, Phys. Rev. B **47**, 7501 (1993); P. Day, M. Lysek, M. LaMadrid, and D. Goodstein, *ibid.* **47**, 10716 (1993); H.S. Youn, X.F. Meng, and G.B. Hess, *ibid.* **48**, 14556 (1993).
[5] S. Prestipino, G. Santoro, and E. Tosatti, Phys. Rev. Lett. **75**, 4468 (1995).
[6] P.B. Weichman and A. Prasad, Phys. Rev. Lett. **76**, 2322 (1996).
[7] S. Prestipino and E. Tosatti, Phys. Rev. B **59**, 3108 (1999).
[8] E.A. Jagla, S. Prestipino, and E. Tosatti, Phys. Rev. Lett. **83**, 2753 (1999); E.A. Jagla and E. Tosatti, Phys. Rev. B **62**, 16146 (2000).
[9] S. Prestipino and E. Tosatti, Phys. Rev. B **57**, 10157 (1998).
[10] G. Trimarchi and S. Prestipino (unpublished).
[11] T. Ohta and K. Kawasaki, Prog. Theor. Phys. **60**, 365 (1978).
[12] S. Prestipino and E. Tosatti, Surf. Sci. **377-379**, 509 (1997).
[13] W. Selke and A.M. Szpilka, Z. Phys. B **62**, 381 (1986).
[14] S. Prestipino and E. Tosatti, Philos. Mag. B **81**, 637 (2001).
[15] J.-R. Lee and S. Teitel, Phys. Rev. B **46**, 3247 (1992).
[16] P.V. Giaquinta and G. Giunta, Physica A **187**, 145 (1992).
[17] See, e.g., F. Saija, S. Prestipino, and P.V. Giaquinta, J. Chem. Phys. **115**, 7586 (2001), and references therein.
[18] S. Prestipino and P.V. Giaquinta, J. Stat. Phys. **96**, 135 (1999); **98**, 507 (2000).
[19] M.G. Donato, S. Prestipino, and P.V. Giaquinta, Eur. Phys. J. B **11**, 621 (1999).